\RequirePackage{fix-cm}
\PassOptionsToPackage{table}{xcolor}
\documentclass[smallextended]{svjour3}       
\smartqed 
\usepackage{amsmath} 
\usepackage{adjustbox} 
\usepackage{cite} 

\usepackage[utf8]{inputenc} 

\usepackage{graphicx}
\usepackage[table]{xcolor}
\usepackage{amsmath}

\AtBeginDocument{%
  \providecommand\BibTeX{{%
    \normalfont B\kern-0.5em{\scshape i\kern-0.25em b}\kern-0.8em\TeX}}}

\usepackage{balance}  
\usepackage{graphics} 
\newcommand{\gias}[1]{\noindent\textcolor{red}{[Gias: {#1}}]}

\usepackage{adjustbox}

\usepackage{url}
\usepackage{fancybox}
\usepackage{multirow}
\usepackage{flushend}
\usepackage{booktabs}
\usepackage{tabularx}
\usepackage{comment}
\usepackage{array}
\usepackage[flushleft]{threeparttable}
\usepackage{mdframed}
\graphicspath{{}{images/}{dia/}}
\DeclareGraphicsExtensions{.pdf,.png}

\usepackage{listings}

\usepackage{courier}
\usepackage{hyperref}

\usepackage{xpatch}

\xpatchcmd{\refstepcounter}{%
  \stepcounter{#1}%
}{%
  \stepcounter{#1}%
  \xdef\scr{\number\value{#1}}%
}{\typeout{success}}{\typeout{failure}}

\newcounter{o}
\usepackage{soul}
\usepackage{tikz}
\definecolor{1c1}{RGB}{188,162,6}
\definecolor{1c2}{RGB}{137,129,80}
\definecolor{1c3}{RGB}{239,167,31}
\definecolor{1c4}{RGB}{88,194,241}
\definecolor{1c5}{RGB}{6,180,188}

\tikzset{mynode/.style={draw=white,solid,circle,fill=green,inner sep=1pt, thick,
text=black}}
\tikzset{arrow line/.style={dashed, line width= 2.5pt, color=#1}}

\usepackage{tcolorbox}

\tcbuselibrary{skins}
\newtcolorbox{rqbox}[1]
{
  before skip=1em, 
  after skip=1em, 
  colframe = black!60,
  colback  = black!10,
  coltitle = black!90,  
  title    = \textbf{#1},
  hbox boxed title,
  enhanced,
  attach boxed title to top center={yshift=-3mm,yshifttext=-1mm},
  boxed title style={size=small, colback=black!30}
} 

\def\bf{\textbf}

\def\fig {Figure~}

\def\tbl {Table~}

\def\sec {Section~}

\def\it{\textit}

\def\tt{\mct}

\newcommand{\mct}[1]{{\footnotesize {\texttt {#1}}}}

\usepackage{paralist}

\newcommand{\nd}{\vspace{1mm}\noindent}
\usepackage{caption}
\usepackage{tikz}

\lstdefinestyle{inlinecode}{basicstyle={\ttfamily\scriptsize\bfseries}}

\newcommand{\urls}[1]{{\scriptsize\url{#1}}}
\usepackage{tcolorbox}
\newcommand{\emt}[1]{\emph{``#1''}}

\usepackage{paralist}
\usepackage[outercaption]{sidecap}
\usepackage [autostyle, english = american]{csquotes}
\newcounter{scn}
\usepackage{tikz}
\usetikzlibrary{positioning,shadows}

\newif\ifpienumberinlegend
\pgfkeys{/number in legend/.code=
    \expandafter\let\expandafter\ifpienumberinlegend
    \csname if#1\endcsname
    \ifpienumberinlegend

    \def\beforenumber##1\afternumber{}%
    \fi,
    /number in legend/.default=true
}
\definecolor{1c1}{RGB}{188,162,6}
\definecolor{1c2}{RGB}{137,129,80}
\definecolor{1c3}{RGB}{239,167,31}
\definecolor{1c4}{RGB}{88,194,241}
\definecolor{1c5}{RGB}{6,180,188}

\tikzset{mynode/.style={draw=white,solid,circle,fill=green,inner sep=1pt, thick,
text=black}}
\tikzset{arrow line/.style={dashed, line width= 2.5pt, color=#1}}

\usepackage{bchart}

\definecolor{ao(english)}{rgb}{0.0, 0.5, 0.0}
\def\test#1{%
\ifnum0#1>0
      #1\%
    \fi
}

\let\oldquote\quote
\let\endoldquote\endquote
\renewenvironment{quote}[2][]
  {\if\relax\detokenize{#1}\relax
     \def\quoteauthor{#2}%
   \else
     \def\quoteauthor{#2~---~#1}%
   \fi
   \oldquote}
  {\par\nobreak\smallskip\hfill(\quoteauthor)%
   \endoldquote\addvspace{\bigskipamount}}
\usepackage[noorphans,vskip=0.5ex]{quoting}

\hypersetup{
    colorlinks=true,
    linkcolor=blue,
    filecolor=magenta,      
    urlcolor=cyan,
}

\usepackage[outercaption]{sidecap}    
\usepackage[figuresright]{rotating}
\usepackage{booktabs}
\usepackage{soul}
\usepackage{subfigure}
\usepackage{pgfplots}
\usepackage{pgfplotstable}

\usepackage{balance}  
\usepackage{graphics} 
\usepackage{color}
\usepackage{booktabs}
\usepackage{ccicons}

\usepackage{url}
\usepackage{fancybox}
\usepackage{multirow}
\usepackage{flushend}
\usepackage{booktabs}
\usepackage{tabularx}
\usepackage{comment}
\usepackage{array}
\usepackage[flushleft]{threeparttable}
\usepackage{mdframed}
\graphicspath{{}{images/}{dia/}}
\DeclareGraphicsExtensions{.pdf,.png,.svg}

\usepackage{listings}
\usepackage{courier}
\usepackage{hyperref}
\usepackage{xpatch}

\xpatchcmd{\refstepcounter}{%
  \stepcounter{#1}%
}{%
  \stepcounter{#1}%
  \xdef\scr{\number\value{#1}}%
}{\typeout{success}}{\typeout{failure}}

\usepackage{tikz}
\definecolor{1c1}{RGB}{188,162,6}
\definecolor{1c2}{RGB}{137,129,80}
\definecolor{1c3}{RGB}{239,167,31}
\definecolor{1c4}{RGB}{88,194,241}
\definecolor{1c5}{RGB}{6,180,188}

\usepackage{listings}

\tikzset{mynode/.style={draw=white,solid,circle,fill=green,inner sep=1pt, thick,
text=black}}
\tikzset{arrow line/.style={dashed, line width= 2.5pt, color=#1}}
\usepackage{varwidth}
\def\bf{\textbf}

\def\fig {Figure~}

\def\tbl {Table~}

\def\sec {Section~}

\def\it{\textit}

\def\tt{\mct}

\usepackage{paralist}

\usepackage{tikz}

\lstdefinestyle{inlinecode}{basicstyle={\ttfamily\scriptsize\bfseries}}

\usepackage{tcolorbox}

\usepackage{paralist}
\usepackage[outercaption]{sidecap}
\usepackage{caption}
\usepackage [autostyle, english = american]{csquotes}

\usepackage[shortlabels]{enumitem}
\usepackage{tikz}
\usetikzlibrary{trees,positioning,shapes,shadows,arrows}
\usepackage[edges]{forest}
\usepackage{pgf-pie}
\usepackage{paralist}
\usepackage{soul}
\usepackage[utf8]{inputenc}
\usepackage[english]{babel}
\usepackage{float}
\usetikzlibrary{positioning}

\usetikzlibrary{positioning,shadows}
\usepackage{balance}

\newif\ifpienumberinlegend
\pgfkeys{/number in legend/.code=
    \expandafter\let\expandafter\ifpienumberinlegend
    \csname if#1\endcsname
    \ifpienumberinlegend

    \def\beforenumber##1\afternumber{}%
    \fi,
    /number in legend/.default=true
}
\definecolor{1c1}{RGB}{188,162,6}
\definecolor{1c2}{RGB}{137,129,80}
\definecolor{1c3}{RGB}{239,167,31}
\definecolor{1c4}{RGB}{88,194,241}
\definecolor{1c5}{RGB}{6,180,188}

\usepackage{tcolorbox}
\tcbuselibrary{listings,skins}
\tikzset{mynode/.style={draw=white,solid,circle,fill=green,inner sep=1pt, thick,
text=black}}
\tikzset{arrow line/.style={dashed, line width= 2.5pt, color=#1}}

\usepackage{bchart}
\usepackage{xcolor}

\tikzstyle{chart}=
[legend label/.style={font={\small},anchor=west,align=left},
legend box/.style={rectangle, draw, minimum size=5pt},
axis/.style={black,thin,->},
axis label/.style={anchor=east,font={\tiny}}]

\tikzstyle{bar chart}=[
chart, bar width/.code={
    \pgfmathparse{##1/2}
    \global\let\bar@w\pgfmathresult},
bar/.style={thin, draw=black},
bar label/.style={font={\bf\small},anchor=north},
bar value/.style={font={\footnotesize}},
bar width=.75,]

\tikzstyle{pie chart}=
[chart,
slice/.style={line cap=round, line join=round, thin, draw=black},
pie title/.style={font={}},
slice type/.style 2 args={
    ##1/.style={fill=##2},
    values of ##1/.style={}}]

\pgfdeclarelayer{background}
\pgfdeclarelayer{foreground}
\pgfsetlayers{background,main,foreground}

\newcommand{\tempsum}{0}

\newcommand{\pye}[3][]{
    \begin{scope}[#1]
        \pgfmathsetmacro{\curA}{0}
        \pgfmathsetmacro{\r}{1}
        \def\c{(0,0)}
        \node[pie title] at (270:1.3) {#2};
        \def\tempsum{0}
        \foreach \v/\s in{#3}{
          \pgfmathparse{\v+\tempsum}
          \global\let\tempsum=\pgfmathresult}
        \foreach \v/\s in{#3}{
            \pgfmathsetmacro{\deltaA}{\v/\tempsum*360}
            \pgfmathsetmacro{\nextA}{\curA + \deltaA}
            \pgfmathsetmacro{\midA}{(\curA+\nextA)/2}
            \path[slice,\s] \c
            -- +(\curA:\r)
            arc (\curA:\nextA:\r)
            -- cycle;
            \pgfmathsetmacro{\d}{min(1.2,max((1.5-0.1*\v) , .5)}

            \begin{pgfonlayer}{foreground}
                \path \c -- node[pos=\d,pie values,values of \s]{$\v\%$} +  (\midA:\r);
            \end{pgfonlayer}
            \global\let\curA\nextA}
    \end{scope}}

\def\test#1{%
    \ifnum #1 > 0
      #1
    \fi
}

\newcommand{\lstbg}[3][0pt]{{\fboxsep#1\colorbox{#2}{\strut #3}}}
\lstdefinelanguage{diff}{
  morecomment=[f][\lstbg{red!20}]-,
  morecomment=[f][\lstbg{green!20}]+,
  morecomment=[f][\textit]{@@},
}

\newcommand{\definebox}[2]{%
  \newcounter{#1}
  \newenvironment{#1}[1][]{%
    \stepcounter{#1}%
    \mdfsetup{%
        frametitle={%
            \tikz[baseline=(current bounding box.east),outer sep=0pt]
            \node[anchor=east,rectangle,fill=white]
            {\strut \MakeUppercase#1~\csname the#1\endcsname\ifstrempty{##1}{}{:~##1}};}}%
    \mdfsetup{innertopmargin=0pt,linecolor=#2,%
        linewidth=0.5pt,topline=true,
        frametitleaboveskip=\dimexpr-\ht\strutbox\relax,}%
    \begin{mdframed}[]\relax%
    }{\end{mdframed}}%
}

\definebox{Recommendation}{black}

\usepackage[table]{xcolor}
\usepackage{hyperref}
\usepackage{balance}
\usepackage{tcolorbox}
\usepackage{soul}
\usepackage{listings}
\usepackage{pgfplots}
\pgfplotsset{compat=1.18} 
\usepackage{multirow}
\usepackage{booktabs} 
\usepackage{subcaption} 

\usepackage{fontawesome5}
\usepackage{fancyhdr}
\usepackage{paralist}

\lstdefinestyle{inlinecode}{basicstyle={\ttfamily\scriptsize\bfseries}}

\usepackage{tcolorbox}

\usepackage{paralist}
\usepackage[outercaption]{sidecap}
\usepackage{caption}
\usepackage [autostyle, english = american]{csquotes}

\usepackage{balance}

\newif\ifpienumberinlegend
\pgfkeys{/number in legend/.code=
    \expandafter\let\expandafter\ifpienumberinlegend
    \csname if#1\endcsname
    \ifpienumberinlegend

    \def\beforenumber##1\afternumber{}%
    \fi,
    /number in legend/.default=true
}
\definecolor{1c1}{RGB}{188,162,6}
\definecolor{1c2}{RGB}{137,129,80}
\definecolor{1c3}{RGB}{239,167,31}
\definecolor{1c4}{RGB}{88,194,241}
\definecolor{1c5}{RGB}{6,180,188}

\usepackage{tcolorbox}
\tcbuselibrary{listings,skins}
\tikzset{mynode/.style={draw=white,solid,circle,fill=green,inner sep=1pt, thick,
text=black}}
\tikzset{arrow line/.style={dashed, line width= 2.5pt, color=#1}}
\usepackage{bchart}
\usepackage{xcolor}
\usepackage{framed} 
\usepackage{enumitem}

\tikzstyle{chart}=
[legend label/.style={font={\small},anchor=west,align=left},
legend box/.style={rectangle, draw, minimum size=5pt},
axis/.style={black,thin,->},
axis label/.style={anchor=east,font={\tiny}}]

\tikzstyle{bar chart}=[
chart, bar width/.code={
    \pgfmathparse{##1/2}
    \global\let\bar@w\pgfmathresult},
bar/.style={thin, draw=black},
bar label/.style={font={\bf\small},anchor=north},
bar value/.style={font={\footnotesize}},
bar width=.75,]

\tikzstyle{pie chart}=
[chart,
slice/.style={line cap=round, line join=round, thin, draw=black},
pie title/.style={font={}},
slice type/.style 2 args={
    ##1/.style={fill=##2},
    values of ##1/.style={}}]

\pgfdeclarelayer{background}
\pgfdeclarelayer{foreground}
\pgfsetlayers{background,main,foreground}

\def\test#1{%
    \ifnum #1 > 0
      #1
    \fi
}

\lstdefinelanguage{diff}{
  morecomment=[f][\lstbg{red!20}]-,
  morecomment=[f][\lstbg{green!20}]+,
  morecomment=[f][\textit]{@@},
}

\newcommand{\customTwoBarChart}[8]{%
\pgfmathsetmacro{\resultxmin}{#2 + #3}
\pgfmathsetmacro{\resultxminnext}{#5 + #6}
\begin{figure}[h]
    \centering
    \hspace{-2cm}
    \begin{subfigure}
        \centering
    \begin{tikzpicture}[scale=0.6]
        \begin{axis}[
            xbar stacked,   
            xmin=\resultxmin,         
            xmax=#1,        
            hide y axis,      
            axis x line*=bottom, 
            width=8cm,        
            height=20mm,      
            y dir=reverse,  
            legend style={at={(0.5,2.65)},anchor=north,legend columns=-1,draw=none}, 
            xticklabels={40\%,20\%,0,20\%,40\%,60\%,80\%}, 
            xticklabel style={anchor=north}, 
            tick align=center,
            bar width = 15pt,
            extra x ticks={0}, 
            extra x tick style={ 
                grid=major,
                grid style={black},
                tick label style={anchor=north},
            },
        ]
        \addplot+[xbar, fill=lightgray!90!white, draw=lightgray!90!white, mark=none,forget plot] coordinates {(#2,0)};        
        \addplot+[xbar, fill=orange!85!white, draw=orange!85!white, mark=none, forget plot] coordinates {(#3,0)};
        \addplot+[xbar, fill=cyan!95!white, draw=cyan!95!white, mark=none, forget plot] coordinates {
            (#1,0)
        };

        \addlegendimage{orange,fill=orange}
        \addlegendentry{Not Difficult}
        \addlegendimage{lightgray,fill=lightgray}
        \addlegendentry{Neutral}
        \addlegendimage{cyan,fill=cyan}
        \addlegendentry{Difficult}
        \end{axis}
    \end{tikzpicture} 
    \end{subfigure}
\hspace{1.5cm}
    \begin{subfigure}
        \centering
        \begin{tikzpicture}[scale=0.6]
            \begin{axis}[
                xbar stacked,   
                xmin=\resultxminnext,         
                xmax=#4,        
                hide y axis,      
                axis x line*=bottom, 
                width=8cm, 
                height=20mm,
                y dir=reverse,  
                legend style={at={(0.5,2.65)},anchor=north,legend columns=-1,draw=none}, 
                xticklabels={38\%,20\%,0,20\%,40\%,60\%,80\%}, 
                xticklabel style={anchor=north}, 
                tick align=center,
                ytick pos=left,
                bar width = 15pt,
                extra x ticks={0}, 
                extra x tick style={ 
                    grid=major,
                    grid style={black},
                    tick label style={anchor=north},
                },
            ]

            \addplot+[xbar, fill=yellow!90!black, draw=yellow!90!black, mark=none, forget plot] coordinates {
                (#5,0)
                
            };
            
            \addplot+[xbar, fill=red!90, draw=red!90, mark=none, forget plot] coordinates {
               (#6,0)
            };

            \addplot+[xbar, fill=green!70!black, draw=green!70!black, mark=none, forget plot] coordinates {
                (#4,0)
            };

            \addlegendimage{red,fill=red}
            \addlegendentry{Not Useful}
            \addlegendimage{yellow,fill=yellow}
            \addlegendentry{Neutral}
            \addlegendimage{green,fill=green}
            \addlegendentry{Useful}
            \end{axis}
        \end{tikzpicture}
    \end{subfigure}
    \vspace{-0.5cm}
   \caption{(a) Manual analysis difficulty Perception (b) iSum usefulness perception for #7}
   \label{#8}
   \vspace{-0.1cm}
\end{figure}
}

\newcommand{\customTwoBarChartIsumAnother}[8]{%
\pgfmathsetmacro{\resultxmin}{#2 + #3}
\pgfmathsetmacro{\resultxminnext}{#5 + #6}
\begin{figure}[h]
    \centering
    \begin{subfigure}
        \centering
    \begin{tikzpicture}[scale=0.6]
        \begin{axis}[
            xbar stacked,   
            xmin=\resultxmin,         
            xmax=#1,        
            hide y axis,      
            axis x line*=bottom, 
            width=8cm,        
            height=20mm,      
            y dir=reverse,  
            legend style={at={(0.5,2.65)},anchor=north,legend columns=-1,draw=none}, 
            xticklabels={20\%,0,20\%,40\%,60\%,80\%,95\%}, 
            xticklabel style={anchor=north}, 
            tick align=center,
            bar width = 15pt,
            extra x ticks={0}, 
            extra x tick style={ 
                grid=major,
                grid style={black},
                tick label style={anchor=north},
            },
        ]
        \addplot+[xbar, fill=lightgray!90!white, draw=lightgray!90!white, mark=none,forget plot] coordinates {(#2,0)};        
        \addplot+[xbar, fill=orange!85!white, draw=orange!85!white, mark=none, forget plot] coordinates {(#3,0)};
        \addplot+[xbar, fill=cyan!95!white, draw=cyan!95!white, mark=none, forget plot] coordinates {
            (#1,0)
        };

        \addlegendimage{orange,fill=orange}
        \addlegendentry{Not Difficult}
        \addlegendimage{lightgray,fill=lightgray}
        \addlegendentry{Neutral}
        \addlegendimage{cyan,fill=cyan}
        \addlegendentry{Difficult}
        \end{axis}
    \end{tikzpicture} 
    \end{subfigure}
    \begin{subfigure}
        \centering
        \begin{tikzpicture}[scale=0.6]
            \begin{axis}[
                xbar stacked,   
                xmin=\resultxminnext,         
                xmax=#4,        
                hide y axis,      
                axis x line*=bottom, 
                width=8cm, 
                height=20mm,
                y dir=reverse,  
                legend style={at={(0.5,2.65)},anchor=north,legend columns=-1,draw=none}, 
                xticklabels={80\%,60\%,40\%,20\%,0,20\%,40\%}, 
                xticklabel style={anchor=north}, 
                tick align=center,
                ytick pos=left,
                bar width = 15pt,
                extra x ticks={0}, 
                extra x tick style={ 
                    grid=major,
                    grid style={black},
                    tick label style={anchor=north},
                },
            ]

            \addplot+[xbar, fill=yellow!90!black, draw=yellow!90!black, mark=none, forget plot] coordinates {
                (#5,0)
                
            };
            
            \addplot+[xbar, fill=red!90, draw=red!90, mark=none, forget plot] coordinates {
               (#6,0)
            };

            \addplot+[xbar, fill=green!70!black, draw=green!70!black, mark=none, forget plot] coordinates {
                (#4,0)
            };

            \addlegendimage{red,fill=red}
            \addlegendentry{Not Useful}
            \addlegendimage{yellow,fill=yellow}
            \addlegendentry{Neutral}
            \addlegendimage{green,fill=green}
            \addlegendentry{Useful}
            \end{axis}
        \end{tikzpicture}
    \end{subfigure}
   \caption{(a) Manual analysis difficulty perception (b) iSum usefulness perception for #7}
   \label{#8}
   \vspace{-0.1cm}
\end{figure}
}

\begin{document}

\title{Automatic Assistance to Reduce Information Overload during Software Issue Report Understanding}
\author{Salma Begum Tamanna \and
        Gias Uddin \and
        Junaed Younus Khan \and
        Lan Xia \and
        Longyu Zhang
}
\institute{Salma Begum Tamanna \at
              University of Calgary \\
              \email{tamanna.sb13@gmail.com}          
               \and
                Gias Uddin \at
               York University \\
               \email{guddin@yorku.ca}    
                \and
                Junaed Younus Khan \at
               Bangladesh University of Engineering \& Technology \\
               \email{junaed@cse.buet.ac.bd}   
               \and
               Lan Xia \and Longyu Zhang\at
               IBM Canada\\
               \email{lan\_xia@ca.ibm.com, longyu.zhang@ibm.com}        
}

\date{Received: date / Accepted: date}

\maketitle

\begin{abstract}
Modern issue tracking systems are used to log an issue and track it from its inception to closure. Issue reports with diverse discussions from stakeholders can be large and verbose, making them challenging to analyze quickly. Our developer interviews revealed that they struggled with the diverse information in issue reports and demanded automated techniques to classify and summarize information. We designed a tool, \it{iSum} (issue summarizer), that helps with the automatic classification and summarization of information types discussed in issue reports. The summarizer offers two views: 
\begin{inparaenum}
\item \it{Developer View} takes an issue report as input and outputs visualized summaries of information in the report.
\item \it{Manager View} takes a suite of all issue reports under a given software project as inputs and shows the prevalence and trends of different information types across the reports.
\end{inparaenum} We evaluated iSum through three studies. 
\begin{inparaenum}
\item In an empirical study over two benchmarks, iSum offers, on average, 11-13\% higher F1-scores than state-of-the-art baselines to detect information types.
\item We conducted a user study in task-based setup with 30 developers to determine iSum's effectiveness in issue exploration scenarios. Results showed increased productivity and accuracy with iSum compared to manual inspection.
\item We assessed iSum's feasibility for development and management tasks. It proved useful to both developers and project managers for issue comprehension and trend monitoring.
\end{inparaenum}

\end{abstract}

\section{Introduction}\label{sec:introduction}
Modern issue tracking systems (ITS) are crucial components in software 
development to log an issue/bug/defect about a software product and track the progress of the issue management from its inception to closure.
Issue reports are often the cornerstone for 
team performance observability and monitoring in a software organization \cite{package, predictissuetype, DynContxtIssueLifetime}. 
Various tasks are performed to manage issues in ITS, which can broadly be 
divided into four types: issue understanding, monitoring, triaging, and finally fixing.

A significant body of research in software engineering (SE) has dedicated to utilize issue reports for software requirement analysis \cite{JITRequirement}, to determine software design decisions \cite{DesignTopicDevDiscuss}, to understand the life cycle of bugs and their influence over different SE stakeholders \cite{secretlifeofbugs}, to categorize issue reports  (e.g., features, bugs, or questions) \cite{predictissuetype,bertissuereport,SWFeatureDetectionIssueTriage,transferbug,topicbugreportdetection,hybridbugreportdetection,bugornot,nonbugfrombug,lstmbug,supervisedbug,fuzzybug}, to 
develop tools and techniques for automated issue triaging and fixing \cite{AutomatedBugTriagingTopicModel,triagetoss,ImprovingBT,Li2021RevisitingTF,Zhang2017EnLDAAN,Kardoost2022DevRankerAE,Tao2017BugRE,Xi2018AnEA,Alazzam2020AutomaticBT,Wang2021EffectiveBT} and others. An important prerequisite to support all such activities is issue understanding. 
Issue understanding denotes the activities of stakeholders to understand the contents of an issue. Various types of information exist in an issue report \cite{InfoTypesOSSDiscussion}, such as comments about bug reproduction, solution discussion, etc. However, the abundance of issue reports or the plethora of information within a single large and verbose issue report can lead to information overload, making them difficult to analyze and comprehend effectively. Additionally, it is not clear which information in an issue report  can help in understanding the issue and whether developers need tool support to understand large issues quickly and efficiently.

This paper focuses on helping software developers and team leads during their 
understanding of an issue. The project started as part of a collaboration with an 
industrial team from a large international software company. The team collaborates with OSS developers in several of their 
open-source industrial projects, such as the design and operationalization of runtime systems (e.g., \underline{\href{https://www.eclipse.org/openj9/}{OpenJ9}}). Our discussions with the team find that issue understanding is often challenging, especially when issue reports are quite large and verbose. Due to the conversation among project stakeholders, these discussion threads accumulate a wealth of knowledge over time regarding the particular problem or the project, and stakeholders in the project can use those various pieces of information according to their needs. A triager may seek out interested or possible contributors, developers may wish to find out the issue replication circumstances; or managers may monitor the solution progress in the thread of conversation. However, issue discussion threads can grow long and complex \cite{pullbasedevelopment} over time and overload the discussion with diverse information, making it challenging for project stakeholders to follow and locate the information they're seeking. In light of this, we recognized the necessity of automatic summarization of various information types as a way to support the understanding of issue reports.

We began our study by interviewing 10 software professionals to gain insight into their direct experiences with issue discussion forums. We explored how they interact with issue reports and identified the challenges they face during issue understanding. Additionally, we assessed the demand for automated assistance tools that enable rapid yet efficient understanding of issues and examined the specific informational needs of different stakeholders. Our findings revealed that developers struggle with the diverse information discussed in issue reports and expressed the need for automated techniques to classify and summarize various information types, such as bug reproduction or solution discussions. To address this need, we introduce iSum (issue summarizer), an interactive toolkit that offers visualized summaries of the different information types in issue reports. To generate these summaries, iSum features a suite of classification models that can automatically detect information types in issue reports: \begin{inparaenum} \item Stand-alone: as stand-alone models, we experimented with LightGBM \cite{lightgbm}, BERT, and RoBERTa. Among them, LightGBM surpassed the state-of-the-art (SOTA) model \cite{InfoTypesOSSDiscussion} of Random Forest (RF) by 6-8\% in overall F1 score. \item Ensemble: We then used these three stand-alone models in an ensemble with the baseline model to generate labels and pass them to the ensemble classifier with textual and conversational features. As the ensemble classifier, again, we experimented with Random Forest, LightGBM, BERT, and RoBERTa. The best ensemble model in iSum achieved 11-13\% higher accuracy than the existing SOTA model \cite{InfoTypesOSSDiscussion}.  \end{inparaenum} We released iSum as an open-source GitHub plug-in, making it available for everyday industrial use and for other researchers to analyze information types in issue reports.


In addition to analyzing iSum's performance across two datasets, we conducted a user study to evaluate the effectiveness of iSum in supporting developers during real-world issue understanding scenarios. In a task-based setup involving 30 developers, we observed that developers were more productive and accurate when using iSum to assess issue reports for understanding or fixing bugs, compared to their performance without it. In a subsequent phase, we evaluated the feasibility of deploying iSum for developers and software teams to enhance their daily issue report comprehension and management tasks. We conducted interviews with 10 industrial and OSS developers, providing them with access to iSum while demonstrating its capabilities. We found that iSum was beneficial for both developers and project managers, aiding in understanding individual issue reports and monitoring trends and discussions across project issue reports. To the best of our knowledge, this is the first study to analyze the importance of issue report understanding using information types and assess its feasibility in professional settings by consulting both industrial and OSS developers.


\section{Understanding Automatic Assistance Needs for Issue Report Understanding} \label{sec:interview1}

Past literature has claimed that navigating complex and lengthy issue discussions to find relevant information is challenging, especially for those unfamiliar with the issue's context, and established a correlation between extended discussions, increased stakeholder involvement, and delays in issue resolution \cite{InfoTypesOSSDiscussion,QuerTCI}. They've also observed that redundant questions often pop up, even after they've been answered before in these lengthy issue discussion threads. However, there is a gap in understanding, from a practical perspective, how developers and project managers interact with issues in their everyday work environments without the aid of automated tools.

We conducted a study to address this gap by examining the hands-on experiences of these key stakeholders and complementing the existing body of knowledge by adding a layer of user-centric analysis. We seek to uncover how open source software (OSS) and industry professionals engage with issue discussion forums, their perspectives on the importance of different types of information in these reports, and the challenges they encounter. This interview aims to evaluate the need for automated assistance tools, with the findings helping to customize these tools to better address the specific needs and challenges faced by users. 

\subsection{Study Methodology}\label{sec:interview1studymethodology}
 \begin{table}[!t]
 \caption{Demographic info of interview participants by years in industry, role and geographic location and team size}
 \vspace{-3mm}
    \centering
    \begin{tabular}{lllll} 
     \toprule
     \bf{P\#} & \textbf{\#Yrs} & \bf{Role} & \bf{Location} & \bf{Team Size} \\ 
     \midrule 
     P01 & 2 & Developer & Asia & 8\\ 
     P02 & 15 & Project Manager & Asia & 140 \\ 
     P03 & 4.5 & Developer & Europe & 18\\ 
     P04 & 5 & Developer \&  Project Manager & Asia & 7 \\
     P05 & 3.5 & Developer & Asia & 3\\ 
     P06 & 4 & Developer & Asia & 2\\
     P07 & 4 & Developer & Asia & 35\\
     P08 & 4 & QA Engineer & North America & 50\\
     P09 & 15 & Developer & North America & 6\\
     P10 & 15 & QA Engineer & Asia & 11\\
     \bottomrule
    \end{tabular}
    \label{tab:interviewparticipants}
\end{table} 

A series of semi-structured interviews have been conducted by the first author with 10 professional software engineers. We invited seven software engineers from our personal contacts, and the rest of the respondents were from our industry partner, IBM. These participants are situated in three different countries (Canada, Bangladesh, and Ireland) from three continents, and their roles spread across the spectrum of engineering and leadership roles. Some of our participants were working in large corporations like Amazon, Samsung, and IBM, and one respondent has their own software startup. We have listed the details of the participants in \tbl\ref{tab:interviewparticipants}.
\begin{inparaenum}[(1)]
    \item Among the participants, two have experience of working in the software industry as project manager. Six of the participants were in the developer role only and remaining two are QA Engineers. 
    \item 7 participants have industry experience of 2–5 years, while other three have around 15 years of experience.  
    \item Five respondents work/last worked in a team with less than 10 members. Others have/had a team of 11–140 members.
    \item Among all the interviewees, four experts identified themselves as female and rest are male.
    \item Everyone has experience of using ITS for issue management; though seven members directly use GitHub; others use JIRA or other internal ITS.
\end{inparaenum}

All interviews were carried out and recorded via Microsoft Teams to facilitate transcription. Due to the limitations of auto-transcription accuracy, each interview underwent manual correction. Following the collection of demographic information, participants were asked to provide feedback on predefined interview questions, all of which were closed-ended. The response options were as follows: Q1–Q5 were binary (Yes/No), and Q6 allowed multi-selection from 16 predefined information types. \tbl\ref{tab:interviewprestudy} presents the interview questions.

 \begin{table}[!t]
\caption{Interview Questions: Understanding Issue Report Exploration Approaches and Challenges}
 \vspace{-3mm}
    \centering
    \begin{tabular}{c p{70mm} l}
     \toprule
     \bf{Q\#} & \bf{Questions} & \bf{Topic}\\ 
     \midrule 
     1 & Do you go through the issue discussion along with the issue details after the assignment? & Process\\ 
     2 & Do you keep yourself up to date with the issue discussion in ITS for the assigned issues? & Process\\ 
     3 & Do you often find that issue discussions contain some necessary information? & Process\\ 
     4 & Have you faced troubles in finding information from lengthy or complex issue threads? & Challenge\\
     5 & Do you think automated info detection from issue discussion can help stakeholders? & Assistance Needs\\
     6 & What type of information do you mostly search for? & Assistance Needs\\
     
     \bottomrule
    \end{tabular}
    \label{tab:interviewprestudy}
\end{table}

\subsection{Issue Report Understanding Processes} \label{sec:understanding} Participants aimed to read an issue report contents and to monitor ongoing discussions in the report during their fixing of a bug or the implementation of a feature as denoted in the report.

\subsubsection{Reading}\label{sec:reading}
{80\% \it{Participants usually read the whole issue thread}~\color{black!90}\rule{80pt}{6pt} }\\
{10\% \it{Participants sometimes skip issue description } ~\color{black!50}\rule{10pt}{6pt} }\\
{10\% \it{Participants sometimes skip issue discussion}~\color{black!20}\rule{10pt}{6pt} }\\
The typical approach among developers when they are assigned an issue is to read the entire issue thread comprehensively to understand the context and specifics of the issue. This method is commonly adopted by 80\% of participants, indicating that they examine the entire discussion when they receive an issue. However, there are instances where more experienced developers might deviate from this approach. For instance, two participants mentioned that they might skip parts of the issue report—such as the description or the discussion—if they feel they already have a sufficient understanding of the issue from the other details provided. Participant P03 noted, \textit{ "I used to see all the technical details and discussion forum when I was new, but after some years, I see the issue title first, and if it's meaningful and it seems good to me that the issue title is enough for me, then I don't go to all the technical details"} - P03.  Additionally, those who have roles in project management tend to briefly review new issues to understand the project's current status but do not engage deeply with the content unless necessary.

Despite varying degrees of engagement with the content of the issues, the consensus among all participants is that issue discussions can be a valuable resource. As P02 emphasized, \it{ "If anyone is interested to know what's happening in this issue, they should go through each line of the issue"} - P02. 

\noindent {100\% \it{Participants consider issue discussions a valuable resource of information}~\color{black!90}\rule{100pt}{6pt} } \\

\subsubsection{Monitoring}\label{sec:monitoring}

{80\% \it{Participants always stay up to date with issue discussion}~\color{black!90}\rule{48pt}{6pt} } \\
{10\% \it{Participants check updates occasionally } ~\color{black!50}\rule{6pt}{6pt} } \\
Nearly all participants (80\%) follow updates on issues they are involved with. This proactive approach is crucial for keeping track of the resolution process, as highlighted by Participant P04, who stated, \it{ "That's necessary, because that way we have to keep track of everything that's been going on"} - P04. Two participants even subscribe to email notifications to ensure they don't miss any updates. In smaller teams, the dynamic changes slightly, with developers often preferring direct communication over discussing issues through information tracking systems (ITS).

\begin{tcolorbox}[flushleft upper, boxrule=1pt, arc=0pt, left=0pt, right=0pt, top=0pt, bottom=0pt, colback=white, after=\ignorespacesafterend\par\noindent]
\textbf{Summary of Processes to Understand Issue Report.} The usual practice of the developers is to go through the whole discussion thread of their assigned issues and keep up to date with the ongoing issue discussion since oftentimes they find valuable information from the discussion.\end{tcolorbox}


\subsection{Issue Report Understanding Challenges} \label{sec:challenges}
{80\% \it{Participants frequently faced difficulties in issue understanding}~\color{black!90}\rule{48pt}{6pt}} \\
{10\% \it{Participants rarely faced difficulties in issue understanding} ~\color{black!50}\rule{6pt}{6pt}} \\
\noindent When exploring the experiences of participants with navigating lengthy and complex issue threads,the feedback highlighted a common challenge: difficulties in extracting necessary information from these extended conversations. This difficulty was echoed by nearly 80\% of the respondents. One participant, P04, described the frustration: \it{"It actually happens quite frequently. Sometimes I go through the whole forum, but don't get any results, or there is no clear answer that it has solved or not"} - P04. This statement captures the essence of the challenge faced by many developers and project managers—diving through extensive discussions without guarantee of finding the information they need. Contrarily, participant P09 had a different experience, stating that such difficulties occurred only rarely. This variation in experience points to the subjective nature of the issue, possibly influenced by the individual's familiarity with the project or their ability to efficiently navigate through discussions. P10 shared insights into the practical approach taken when faced with voluminous threads; repeatedly reading through dense material to ensure no critical information is overlooked: \it{"All comments may not be necessary but I don't know which is important and which is not. So, I need to go over everything, and when I'm going through all the comments, it's possible that while I'm just glancing, I might have overlooked the important points. Or sometimes I get lost in reading such lengthy things and I repeat reading it"} - P10. This response illustrates the depth of the challenge, where even seasoned professionals find themselves re-reading and scrutinizing discussions to ensure no critical information is missed.

\begin{tcolorbox}[flushleft upper, boxrule=1pt, arc=0pt, left=0pt, right=0pt, top=0pt, bottom=0pt, colback=white, after=\ignorespacesafterend\par\noindent]
\textbf{Summary of Challenges during Understanding Issue Report.} The primary challenge identified by participants in understanding issue reports was the difficulty of extracting essential information from long and complex discussion threads. Nearly 80\% struggled with this often and finding it hard to locate the required information. This led to repetitive reading and scanning of conversations, with some participants worrying they might miss critical details.
\end{tcolorbox}

\begin{table}[!h]
\rowcolors{2}{}{lightgray!30}
\caption{Definition of information types found on issue discussion (Content Mgt = Issue Content Management)}
\begin{adjustbox}{width=\columnwidth}
    \centering
\fontsize{8pt}{8pt}\selectfont
    \begin{tabular} {p{22mm}|p{42mm}|p{42mm}}
    \toprule
        \textbf{Info Type} & \textbf{Definition (based on~\cite{InfoTypesOSSDiscussion})} & \textbf{Examples (from OpenJ9 Issue Reports)} \\ \midrule
        Action on Issue & Comments about actions perform for issue management like closing, reopening, labeling and others  & 
        \emt{Closing, since Java 11 Windows is now compiling with MSVC 2017.} \\ 
        Bug Reproduction & Any question, statement or report regarding the reproduction of the bug 
        & \emt{I still cannot reproduce this issue.} \\ 
        Contribution and Commitment & About assigning or asking members for resolving an issue along with volunteer contribution or unwillingness 
        & \emt{please assign to me.}\\ 
        Expected Behaviour & The expected/ideal situation from user's perspective or personal preference and opinion 
        & \emt{I expect it's invalid to access the variable outside of holding the mutex, as the value can be modified by other threads.} \\ 
        Future Plan & Any long term plan related to the issue 
        & \emt{We should design the data record to be extensible in the future.}\\ 
        Investigation and Exploration & Suggesting investigation ideas or reporting results about the cause of the issue  
        & \emt{We need to see the bytecodes to see why the check in the verifier is failing.} \\ 
        Content Mgt & Redirecting issue discussions to another place 
        & \emt{Is it possible to move the discussion to a slack channel?} \\ 
        Motivation & Why an issue needs to resolved or a feature need to be added or changed 
        & \emt{OpenJDK is modifying Windows compilation for Java 
        11+ to use VS2017 and OpenJ9 should support the same.} \\ 
        Observed Bug Behaviour & Questions or reports about the observed behaviour of a bug 
        & \emt{Crash or OutOfMemoryError when using AsyncGetCallTrace()} \\ 
        Potential New Issues and Requests & Comments involving identifying new bugs or need of new features or any change while investigating an issue & 
        \emt{Unfortunately I am now hitting another problem: if I run long enough, the flight service starts consuming all the  CPU on the machine} \\ 
        Social Conversation & Any kind of small talk or  showing gratitude or disappointment or other emotions.   
        & \emt{Thanks for looking into this!} \\ 
        Solution Discussion &  Suggestion or statement about design, implementation, constraints, challenges and references for solution 
        & \emt{I think the 
        assert should just be removed.} \\ 
        Testing &  Any question, statements or report about testing process, environment and results 
        & \emt{Did something change with the SharedClasses.SCM23.MultiCL\_0 test?}\\ 
        Task Progress & Asking or providing update about the progress of the task towards the solution  
        & \emt{Any update on this?} \\ 
        Solution Usage &  Question or suggestion on how to use the new solution. 
        & \emt{Is that not the official way to use BN?}\\ 
        Workarounds & Temporary solution before the official one & \#10848 provides 
        a \emt{temporary fix for the 0.23 release.} \\ \bottomrule
    \end{tabular}
\end{adjustbox}
    
    \label{tab:definition}
\end{table}
\subsection{Needs for automatic assistance for information understanding} \label{sec:studyassistance} 
The participants wished for automatic assistance to automatically detect and summarize various information types discussed in issue reports (eg., bug reproduction, solution discussion, etc.).

\noindent {100\% \it{Participants support for automation detection and summarization of diverse information}~\color{black!90}\rule{60pt}{6pt}} \\
\subsubsection{Relevant Information Identification} \label{sec:infoidentification} 
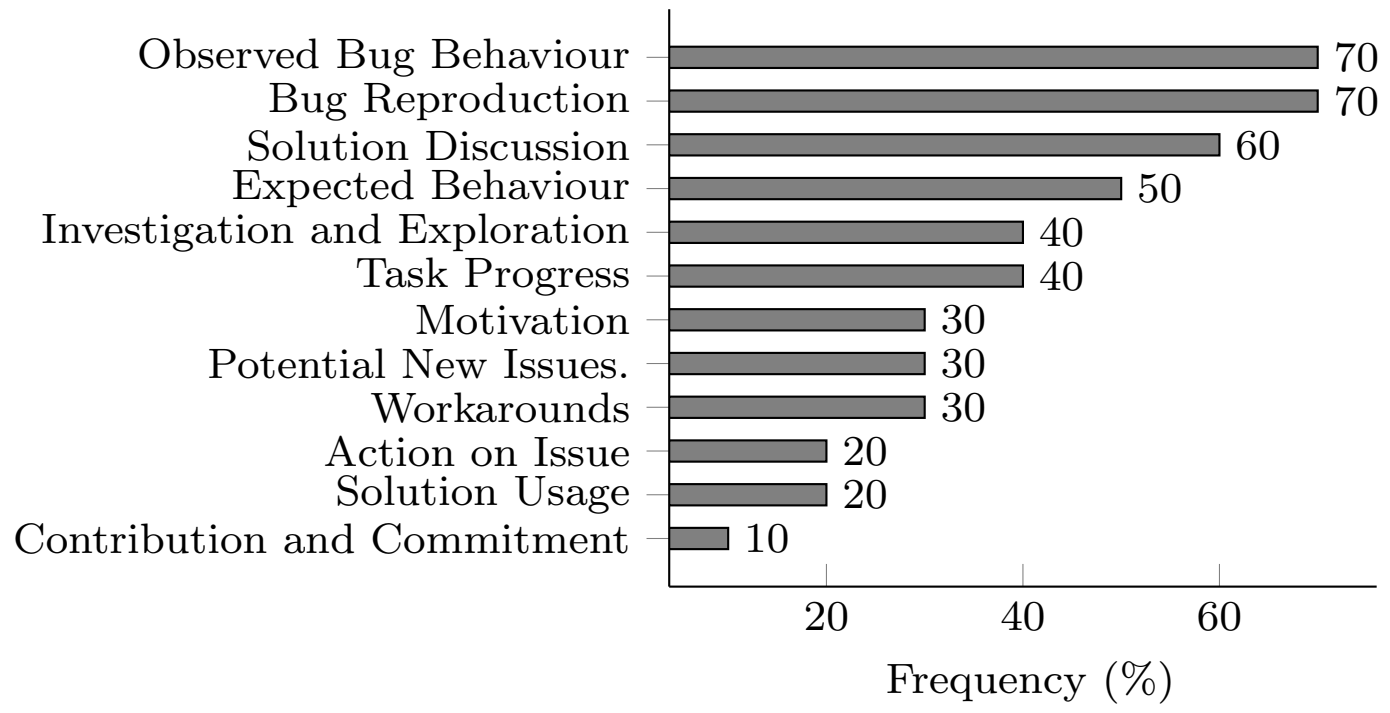
\begin{figure}[!t]
    \centering
\begin{tikzpicture}  
\begin{axis}  
[   axis x line*=bottom,
     axis y line*=left,
     height = 54mm,
    xbar,  
    xlabel={\ Frequency (\%)}, 
    bar width=4pt,
    symbolic y coords={ Contribution and Commitment, Solution Usage,Action on Issue, Workarounds, Potential New Issues.,Motivation,Task Progress,Investigation and Exploration,Expected Behaviour, Solution Discussion,Bug Reproduction,Observed Bug Behaviour},
    ytick=data,  
    nodes near coords, 
    nodes near coords align={horizontal},  
    ]  
\addplot [fill=gray,
        ]coordinates {(20,Action on Issue) (70,Bug Reproduction) (10,Contribution and Commitment) (50,Expected Behaviour) (40,Investigation and Exploration)
(30,Motivation) (70,Observed Bug Behaviour) (30,Potential New Issues.) (60,Solution Discussion) (40,Task Progress) (30,Workarounds) (20,Solution Usage)
};  
\end{axis} 
\end{tikzpicture}  
\caption{Importance of information types to the participants}
\label{fig:infotypeinterview}
\end{figure}

To further understand which information is relevant to stakeholders in their work, we provide them with a list of information types, drawn from the taxonomy built by Arya et al. \cite{InfoTypesOSSDiscussion}. These information types are listed in \tbl\ref{tab:definition} with definitions and examples. This list offers a consistent framework that helps stakeholders quickly identify and select relevant categories, organizing their thoughts more effectively. It also allows us to arrange information into familiar categories, reducing cognitive load and helping users focus on the most pertinent information. Also, by knowing which information types are most sought after, the tools can be tailored to prioritize and highlight these types. Responses showed that most respondents face bug related issues; therefore, they are most interested in finding Observed Bug Behaviour, Bug Reproduction,  Solution Discussion and Expected Behaviour related comments; but for feature enhancement issues, they look for motivation. One respondent said, in case where a quick solution is needed, like when the module owner is not available, they sometimes look for Workarounds. Though some inform for task progress, they use other management software; some still search the comment section to learn about Task Progress. However, no one is interested in reading the Social Conversation. Rather, one suggested to categorize Social Conversation related comments to identify the expressions or toxicity. \fig\ref{fig:infotypeinterview} shows the distribution of the information types that participants find important.

\subsubsection{Relevant information Summarization} 
When asked about the potential benefits of automated assistance for issue summarization, all participants voiced strong support for such technology. The consensus highlights a widespread challenge in current practices—manually navigating through extensive threads is not only time-consuming but also prone to errors, as critical details may be missed or overlooked. Participant P10 expressed the urgency for automation by stating, \it{"I may overlook the important points and I may get lost reading through and I will have to repeat reading, which is time consuming. So anything that is time-consuming can be automated, that is the basic rule of automation"} - P10. This sentiment underscores the common frustration of needing to revisit lengthy threads multiple times to ensure no essential information has been missed. Participant P06 added another dimension to the discussion by emphasizing the need for categorization based on task progress or actions taken by other team members: \it{"I need to know what is the task progress in it or what other people have done on that. So I think if there is anything to distinguish among all the data, that would be better."} This remark illustrates a desire for structured information that can streamline the process of understanding task statuses and historical actions within a project.  All feel this assistance can help all project stakeholders, but two emphasized that it would only help project managers, as commented by P09, \it{" I would have just read the issue, but I think there is a need to categorize it from a project point of view. That's to me, maybe more meaningful"} - P09.

\begin{tcolorbox}[flushleft upper, boxrule=1pt, arc=0pt, left=0pt, right=0pt, top=0pt, bottom=0pt, colback=white, after=\ignorespacesafterend\par\noindent]
\textbf{Summary of Needs for Automatics Assistance for information Understanding.} Due to the information overload in lengthy and complex issues, project stakeholders feel the necessity of automatic identification of issue discussion information types. From the issue discussion, stakeholders mostly check for bug and solution related information.
\end{tcolorbox}


\section{Needs for Information Detection and Summarization in Issue Reports}\label{sec:motivation}
\begin{figure}[!t]
\centering
\hspace*{-0.45cm}
\includegraphics[scale=0.5]{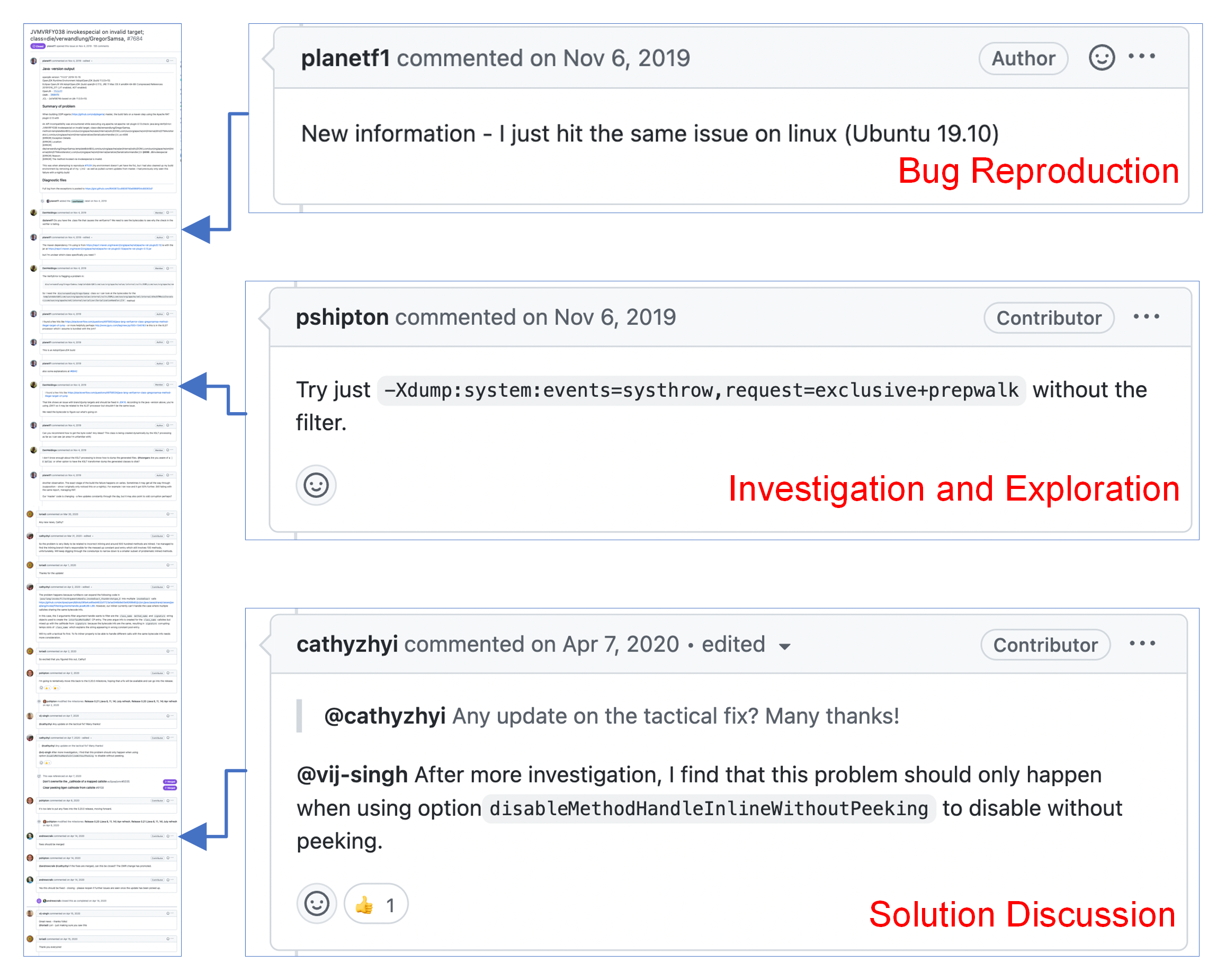}
\caption{Issue discussion thread of OpenJ9 \#7684 where important information are spread in the discussion}
\label{fig:examples}
\end{figure}

The feedback from the interviews demonstrates that issue reports are not just mere notifications of problems; they contain detailed discussions of collaborative problem-solving efforts. Stakeholders rely on these discussions to extract valuable information for issue resolution. We consulted with our industrial partner about the study findings and found that the findings resonated with their difficulties in understanding the issue report. For instance, an OpenJ9 issue thread (refer to Figure \ref{fig:examples}), filled with 155 comments, contains crucial details like bug reproduction steps, suggestions for investigation, and a final solution.  For example, a stakeholder posted a comment stating, \it{"New information—I just hit the same issue on Linux (Ubuntu 19.10)."} This insight could be crucial for understanding the scope of the problem and potentially identifying any platform-specific factors contributing to the issue.  But amidst all these comments, finding the important insights needed to understand the issue can be like searching for a needle in a haystack. The complexity of navigating such dense information is further compounded by the frustrations stakeholders experience when forced to read or re-read lengthy discussions. As encapsulated by Participant P02's remark, \textit{ "when an issue has, for example, 40 comments, sometimes it's annoying to read those comments."} Responses from our interview, presented in Section \ref{sec:challenges}, underscore the inefficiencies and potential for oversight inherent in manual navigation of issue discussions, which not only slows down the resolution process but also increases the risk of missing crucial information.

Recognizing the challenges outlined by participants, there emerges a clear and compelling need for automated tools that can streamline the understanding of issue discussions. During the interviews, all participants responded positively to the idea of automated assistance, particularly the automated identification of information types within issue discussions, indicating a unanimous agreement on its potential benefits. Such tools would mitigate the burdens of manual navigation and enhance the accuracy and efficiency of information retrieval from discussions, as further discussed in Section \ref{sec:studyassistance}.

Prior research, such as that by Arya et al. \cite{InfoTypesOSSDiscussion}, has recognized the importance of categorizing information types within issue discussions. Their motivation stemmed from the understanding that issue reports often contain diverse information types, each serving different stakeholders' needs, from developers seeking bug reproduction steps to project managers looking for task progress updates. Drawing from these insights, it becomes evident that navigating dense, information-rich discussions, as seen in the OpenJ9 project thread, poses a common challenge. Therefore, we focus on developing a more accurate automated assistance that can detect the diverse information from issue discussion and ease issue understanding.

\section{{iSum:} Automatic Summarization of Information Types in Issue Reports}\label{sec:isum}
We have designed iSum (Issue Summarizer), an end-to-end pipeline/framework as a suite of several toolkits that can be used to automatically detect information types discussed in an issue report or across issue reports. The tool iSum then summarizes the detected information types and their associated discussions in interactive and visualized dashboards. 

The design of iSum is motivated by our findings from the interviews (see Section \ref{sec:interview1}) and our exploration of the information overload problems during software issue report understanding, as we undergone these problems with our industrial partners over a span of more than a year (see Section \ref{sec:motivation}). 

In this section, we present the overall framework of iSum and the modules that comprise the framework (see Sections \ref{sec:isum-design-decisions} - \ref{sec:isum-summarization}). We then discuss the implementation architecture of iSum that we deployed as a plug-in in collaboration with our industrial partner and that we used to assess the effectiveness of iSum in real-world tasks. 

\subsection{iSum Framework}\label{sec:isum-design-decisions}
Issue Summarizer (iSum) is a toolkit that outputs visualized summaries of the
various information types present in the issue reports. For this generation of
information type summary, iSum employs a Machine Learning (ML) model that can
automatically identify the information type in each sentence of an issue report. 
We take the catalog of 16 information types from Arya et al.~\cite{InfoTypesOSSDiscussion}.

\begin{figure}[h]
\centering
\includegraphics[scale=0.47]{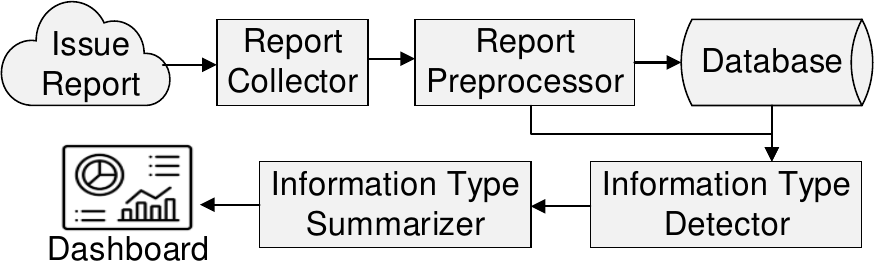}
\caption{Schematic diagram of the iSum framework}
\label{fig:schemaIsum}
\vspace{-3mm}
\end{figure}

The iSum framework has four major components (see \fig\ref{fig:schemaIsum}): two to collect and process issue reports and two to detect and summarize information types in the reports.

\bf{{Report Collector and Preprocessor:}} The collector component is used to download or fetch issue
discussion threads from project repository. The processor  module retrieves total 14 metadata as
conversational features  from the sentences of an issue discussion thread. We
followed the guidelines set by Arya et al.~\cite{InfoTypesOSSDiscussion} for
defining these metadata.  These conversational features include participant
information, sentence length related information, structural (position) and
temporal (time) information and existence of code snippets in a sentence. For
the saved project, sentences and metadata of all issues in the repository get
stored in a database and for a single issue report, these information directly go to the next component.
    
\bf{{Information Type Detector and Summarizer:}} This detector component is used to detect information type for each sentence in an issue report 
(discussed in \sec\ref{sec:model}). The summarizer component provides two dashboards of summaries of information 
types for the issue reports (discussed in \sec\ref{sec:isum-summarization}). 

\subsection{Classification of Information Types} \label{sec:classificationOfInformationTypes}
To classify the types of information within issue discussions effectively, we took a multi-faceted approach. First, we carefully selected four ML models renowned for their distinct strengths: Random Forest, LightGBM, BERT, and RoBERTa. Each model was chosen for its unique capabilities in handling various aspects of the classification task. Random Forest was selected as a baseline following Arya et al.~\cite{InfoTypesOSSDiscussion} because it effectively captures conversational patterns. LightGBM was selected for its ability to efficiently handle high-dimensional sparse TF-IDF features and capture non-linear interactions with conversational features, particularly on small and imbalanced datasets, while providing built-in regularization \cite{lightgbm}. BERT \cite{bert} was chosen for its bidirectional attention mechanism, which captures rich contextual semantics crucial for distinguishing similar information types such as Motivation and Expected Behaviour. Finally, RoBERTa \cite{roberta} was included as it builds upon BERT with improved pre-training strategies, yielding more robust text representations.

Following model selection, we conducted extensive feature engineering to optimize the performance of RF and LightGBM and fine-tuned BERT and RoBERTa. Following this, we employed an ensemble technique to integrate the predictions of these standalone models, enhancing the overall classification accuracy. Notably, we also designated these same four models to serve as the final classifier within our ensemble framework. This decision allows users of iSum to utilize either a single model or an ensemble model according to their specific requirements, ensuring flexibility and adaptability in addressing diverse information classification needs. We will discuss more about the models in section \ref{sec:model} but discuss the feature engineering here.

\subsubsection{Feature Engineering} \label{sec:featureengr}
For feature engineering, we incorporated a range of features that span both textual and non-textual data from issue reports. Depending on the context, our model utilizes either one, both, or none of these features. Prior to feature extraction, we conducted text preprocessing.

\bf{Text Preprocessing.} Initially, we preprocessed the raw text from issue discussions by replacing code snippets with the keyword \it{CODE} and trimming sentences to remove unnecessary whitespace, ensuring cleaner data for analysis.

\bf{Textual Features.}
 For transforming the text into a format suitable for ML models, we utilized a Term Frequency-Inverse Document Frequency (TF-IDF) Vectorizer. This technique converts text data into numerical feature vectors, emphasizing the importance of words based on their frequency across documents but penalized if they appear too commonly across all documents. This helps in highlighting words that are uniquely significant to certain types of discussions in the issue reports. We set \it{max\_features} to 10000 to limit the number of most significant words considered across the documents. This constraint is designed to prevent the model from being overwhelmed with data and to focus on the most informative features, enhancing both efficiency and effectiveness.

\begin{table}[!t]
\centering
\caption{Conversational Features Used in Our Study (CL = Collaborator, norm = normalized)}
\label{tab:metadata}
{\small
\begin{tabular}{lp{1.8cm}p{4.2cm}p{2cm}}
\toprule
\textbf{Category} & \textbf{Feature Name} & \textbf{Description} & \textbf{Value Range} \\ \midrule
\multirow{2}{*}{Participant} & AA & Author's association with the repository & \{Owner, CL, Member, Other\} \\
 & BEGAUTH & Indicates if the author also posted the original issue & \{True, False\} \\
\midrule
\multirow{3}{*}{Length} & LEN & Total character count of the sentence & Positive Numbers \\
 & TLEN & Ratio of sentence length to the longest sentence in the thread & (0, 1] \\
 & CLEN & Ratio of sentence length to the longest sentence in the comment & (0, 1] \\
\midrule
\multirow{4}{*}{Structural} & TLOC & Position of sentence within its comment & (0, 1] \\
 & CLOC & Position of sentence within the entire conversation & (0, 1] \\
 & FIRST\_TURN & Indicates if the sentence is in the first comment & \{True, False\} \\
 & LAST\_TURN & Indicates if the sentence is in the last comment & \{True, False\} \\
\midrule
\multirow{4}{*}{Temporal} & TPOS1 & Time from the start of the conversation to the comment, norm. & [0, 1] \\
 & TPOS2 & Time from the comment to the end of the conversation, norm & [0, 1] \\
 & PPAU & Time from the previous comment to the current comment, norm & [0, 1] \\
 & NPAU & Time from the current comment to the next comment, norm & [0, 1] \\
\midrule
Code & HAS\_CODE & Indicates whether the comment contains a code snippet & \{True, False\} \\
\bottomrule
\end{tabular}}
\end{table}

\bf{Conversational Features.}
To capture the context within which each sentence appears in the discussion, we incorporated a set of conversational features outlined in \cite{InfoTypesOSSDiscussion}, as listed in Table \ref{tab:metadata}:

\begin{enumerate}
\item {Participant Features:} Including the author's association (AA) with the repository and whether the author of the comment also initiated the issue (BEGAUTH).
\item {Length Features:} Such as the total length of the sentence (LEN), the relative length compared to the longest sentence in the thread (TLEN), and within the same comment (CLEN).
\item {Structural Features:} These include the location of the sentence within the comment (TLOC) and within the entire conversation (CLOC), along with flags indicating whether a sentence appears in the first (FIRST\_TURN) or last comment (LAST\_TURN).
\item {Temporal Features:} Measure the timing of a comment within the conversation, such as the time elapsed since the start of the discussion (TPOS1), time until the end of the discussion (TPOS2), and the time intervals between adjacent comments (PPAU and NPAU).
\item {Code Feature:} A binary flag (HAS\_CODE) indicating the presence of a code snippet within a comment, which can be crucial for understanding the technical nature of a discussion.  
\end{enumerate}


\subsection{Summarization of Information Types}\label{sec:isum-summarization}

\begin{figure*}[!t]
\includegraphics[scale=0.23]{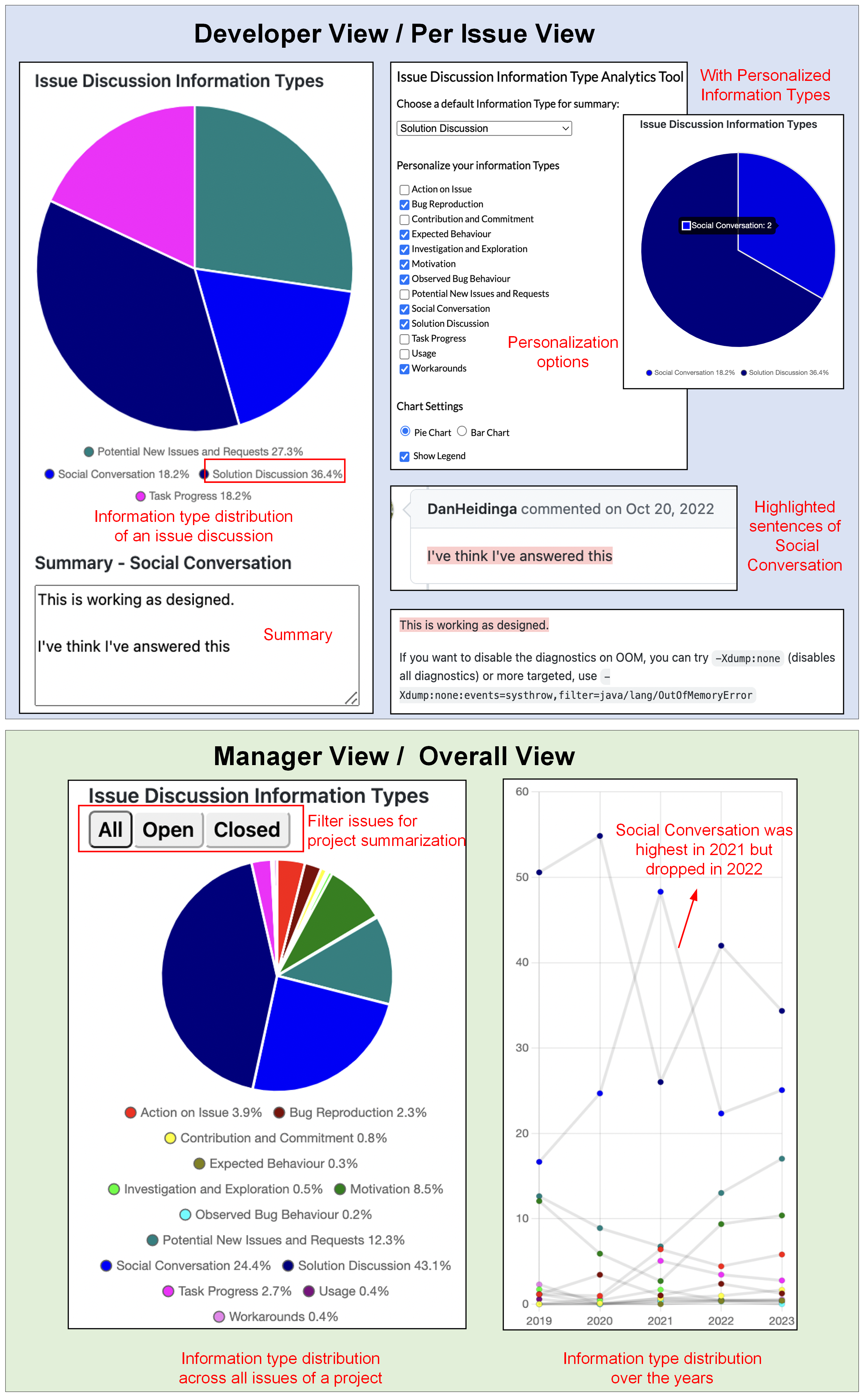}

\caption{The iSum dashboards to summarize information types in developer and manager views}
\label{fig:snippet}
\end{figure*}

The iSum summarizer assists the developers/managers with a suite
of summarization of the detected information types in issue reports. 
The summaries can be of three types, as described below:
\begin{itemize}[leftmargin=10pt]
  \item \bf{Visuals}. We show two types of visuals as follows. 
  \begin{enumerate}
    \item \bf{Distribution.} We show the overall distribution of information
    types at sentence in an issue report and across issue reports. This is
    suitable to get a quick overview of the type
    of discussions happening in an issue or in a project.
    \item \bf{Trends.} We show the distribution of information types in a project. This is suitable to understand whether certain types of discussions are more prevalent in recent days/months compared to before.
  \end{enumerate}
  \item \bf{Text Summaries}. We show a summary of all textual comments for a given information type. The summary is created using the TextRank algorithm~\cite{Mihalcea-TextRank-2004}. This approach enables a quick overview of comments related to a specific information type, as TextRank is a lightweight, unsupervised extractive method that requires no additional training data and performs reliably on short, domain-specific software engineering texts such as issue comments.

  \item \bf{Text Highlighter}. We highlight all comments for a given information type, so that users can check those without reading the entire contents of the issue report. 
\end{itemize}
At this moment, iSum offers two types of dashboards to showcase the summaries for the stakeholders.
\begin{itemize}[leftmargin=10pt]
  \item \bf{Developer View} corresponds to the summaries per issue.  
  \item \bf{Manager View} corresponds to the summaries across all issues.
\end{itemize}
\fig\ref{fig:snippet} shows screenshots of the two dashboards in iSum UI (top: Developer, bottom: Manager View). The views are 
discussed below. 

\subsubsection{Developer View} \label{sec:developerview}
In any issue page, it displays the distribution of information types present in the issue discussion in a pie or bar chart depending on the default settings of the tool.  Tooltip of each slice of the pie chart shows the total number of information type along with name of the information type. Legend maps between the color and name of information type and the current percentage of the distribution. Clicking on any slice of the chart highlights the sentences of the information type and shows the top 5 sentences (if any) as the summary in a textarea below. These top 5 sentences are selected by the TextRank algorithm according to their centrality, that is, the sentences most representative of all sentences of that type, rather than the first 5 or the 5 most recent sentences. This tool also allows users to personalize the visualization. Users can select which information types are displayed in the charts, and the same color is used consistently for both the highlighted sentences and the corresponding chart slice on the GitHub page. Users can also save their preferred settings so that the selected information types persist across sessions.
\subsubsection{Manager View}
At the start page of the saved project repository, iSum displays two charts: \begin{inparaenum}[(1)]
\item the information type distribution over all issue discussions of that
project as pie or bar chart  and \item the evolution of the prevalence of
information types over the years from beginning of the project to current date
with multiple line charts; each corresponds for each information
type\end{inparaenum}. Single legend will map between the color and the
information type of both charts. Users are also able to  filter issue to display
the data about all, open or closed issues using the button displayed.

\subsection{iSum Deployment and Industrial Adoption}\label{sec:isum-architecture}
We built \textit{iSum } as a browser extension tool. The tool is activated when a user goes to a GitHub project that contains issue reports. 
As such, iSum can offer the two dashboards within a GitHub page: the developer view at the per issue level of a project and the manager view at the 
main page of the project in GitHub. 

\begin{figure}[t]
\centering
\includegraphics[scale=0.5]{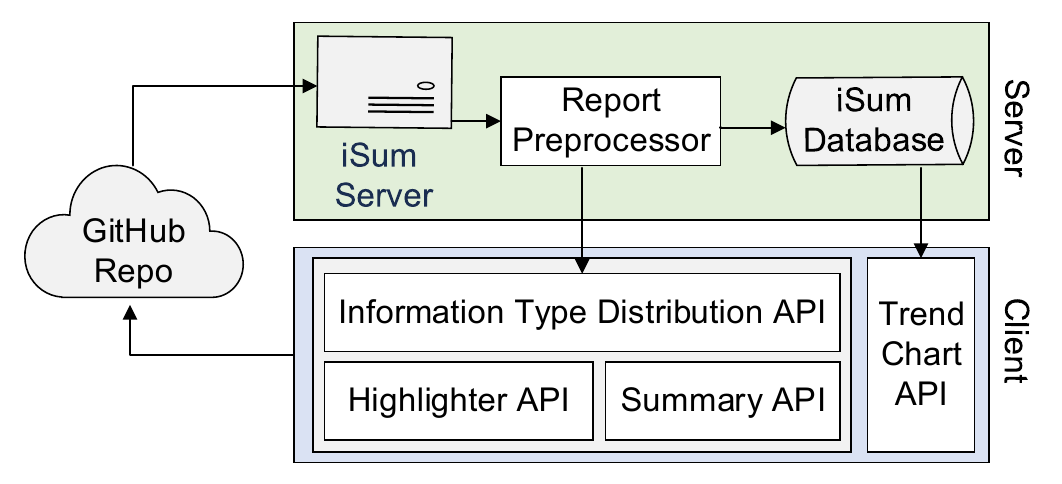}
\caption{The client server architecture of iSum}
\label{fig:client-server}
\end{figure}

To build this extension tool, we have exploited a simple client-server 
architecture that uses a relational database for storing issue reports and is developed on top of the Python Django Web Framework. 
The communication between Django webserver and iSum report processor is handled
via Rest-based middleware to transfer data in JSON format over HTTP protocol. We
do not need an website for output visualization since the UI elements are added
on the GitHub issue/repository page.  Figure \ref{fig:client-server} exhibits
the client-server architecture of iSum. There are three major components of the
iSum: 
\begin{inparaenum}[1)]
    \item {\textit{Report Preprocessor:}} Handles processing of the issue reports and hosts all the techniques to summarize issue discussion information types.
    \item {\textit{Database:}} A data storage component for storing issue discussion sentences with its metadata.
    \item {\textit{REST Web Server:}} The middleware between the Report Preprocessor and the repository.
\end{inparaenum}
Since we follow client server architecture for iSum deployment, the deployment
only needs a server for handling preprocessing of the data and hosting algorithms
and techniques for issue and project summarization. Any OSS user or any
organization that are involved on open-source industrial projects  can easily
deploy the tool in a local server or a server owned by them. %

\section{Performance of Information Type Classification in iSum} \label{sec:model}

Arya et al. \cite{InfoTypesOSSDiscussion} utilized traditional shallow ML models like random forest and logistic regression for classifying information types in issue discussions. However, there is scope for further enhancement in accuracy and efficiency by employing more advanced models and techniques. Enhanced models leveraging advanced ML techniques such as ensemble methods could provide significant improvements. Ensemble methods combine predictions from multiple models to improve accuracy and reliability over individual models. Given the complex and varied nature of issue report data, ensemble models can better capture the nuances and reduce the likelihood of misclassification.  In this section, we answer the following two research questions about the development of these models for information type classification in issue reports and evaluate their performance with two benchmark datasets.

\begin{enumerate}[label=\bf{RQ\arabic{*}.}, leftmargin=1cm]
    \item How do the stand-alone models perform in iSum?
    \item Can we improve it based on an ensemble of models?
\end{enumerate}

\subsection{Studied Datasets} \label{sec:datasets}

We used two benchmark datasets. First, we used the benchmark dataset previously developed by Arya et al. \cite{InfoTypesOSSDiscussion}. 
We refer to this benchmark as Dataset 1. The models developed by Arya et al. \cite{InfoTypesOSSDiscussion} are 
our baselines against which we check our new models. The dataset did not have any contents from the issues of language runtime systems, the software our industrial partner is working on. As such, we also produced another benchmark by analyzing the issues in the runtime systems (OpenJ9). The code and issues of runtime systems are open source and available on \href{https://github.com/tamanna037/ISum}{iSum GitHub repository}. As such, this new benchmark can also be used by the researchers to develop new models for information types. We refer to this benchmark as Dataset 2. In both datasets, a GitHub issue contains a thread—the chronological sequence of comments posted on that issue. Each comment is written by a single participant and  may contain one or more sentences; the sentence is the unit of classification in our work. We use \textit{conversation} and \textit{thread} synonymously to refer to the full 
sequence of comments within a single issue.An issue has multiple threads, each consist-
ing of one or multiple comments, each consisting of one or multiple sentences,

   \begin{table}[t]
   \caption{Dataset used in developed ML models (\#Sen = \# of Sentences, \#Type = \# of Information Types)}
   \vspace{-3mm}
    \centering
    \begin{tabular}{llrr}
     \toprule
    \textbf{Dataset} & \textbf{Source} & \textbf{\#Sen} & \textbf{\#Type}\\ 
     \midrule 

     Dataset 1\cite{InfoTypesOSSDiscussion} & TensorFlow, scikit-learn, Spacy & 4330 & 13\\ 
     Dataset 2 & OpenJ9 & 4241 & 12\\
     \bottomrule
    \end{tabular}
    
    \label{tab:dataset}
\end{table}
\nd\bf{Dataset 1.} We reused the dataset collected by Arya et al. \cite{InfoTypesOSSDiscussion}. It consists of 15 complex issues, taken from
  three Artificial Intelligence (AI) libraries hosted on GitHub, namely, TensorFlow, scikit-learn, and spaCy; 5 issues from each repository. Initially,
  this dataset consists of total 4656 sentences of 16 types of information. However, sentences which have multiple labels are removed along with the sentences labeled with the information types, which have very few sentences associated with them. Future Plan, Content Management and Testing-Related; these three labels have sentences which covers only 1\% of whole dataset and thus removed. So the final dataset has total 4330 sentences containing 13 informaton types.

\nd\bf{Dataset 2.} For this one, we have collected sentences from total 21 closed issues in OpenJ9 GitHub repository. We have chosen this repository since it supports the interest of our industrial partner IBM Canada. The rules set for the selection of issues are similar to the previous work \cite{InfoTypesOSSDiscussion}. Closed issues are chosen to get the full discussion flow of an issue, although these information types are applicable for open issues too. From all the closed issues of OpenJ9, the ones with the most comments up to June 28, 2022 are chosen for preparing this dataset. This dataset has been labeled by the first author of this manuscript, strictly adhering to the label definitions established by Arya et al. \cite{InfoTypesOSSDiscussion}, without any modifications. For this dataset also, labels with few sentences are removed along with the sentences with multilabels. The final dataset contains 4241 sentences of 12 information types, leaving out sentences of the following four labels: Future Plan, Potential New Issues and Requests, Content Management, and Solution Usage.

Dataset-level statistics of the selected issue reports from TensorFlow, scikit-learn, spaCy, and OpenJ9 are summarized in Table \ref{tab:issue_stats}. For each project, we report: (i) the total, mean, and variance of the number of sentences per thread (i.e., thread length in sentences); and (ii) the total, mean, and variance of the number of comments in the dataset.

\begin{table}[!t]
\centering
\caption{Statistics of studied datasets for benchmark projects}
\begin{tabular}{l|l|ccc|ccc}
\toprule
\textbf{Dataset} & \textbf{Project} & \multicolumn{3}{c|}{\textbf{Sentences}} & \multicolumn{3}{c}{\textbf{Comments}} \\
 & & Total & Mean & Var & Total & Mean & Var \\
\midrule
\multirow{3}{*}{Dataset 1} & Scikit-learn & 1401 & 280 & 1759 & 429 & 86 & 338 \\
 & SpaCy        & 829  & 166 & 6587 & 173 & 35 & 148 \\
 & TensorFlow   & 2100 & 420 & 21165 & 751 & 150 & 3092 \\
 \midrule
 Dataset 2 & OpenJ9       & 4241 & 202 & 4194 & 1599 & 76 & 612 \\
\bottomrule
\end{tabular}
\label{tab:issue_stats}
\end{table}

\subsection{Model Validation and Evaluation}
\subsubsection{Validation Approach} \label{sec:validation} 
We used a stratified k-fold cross-validation (k = 10) on a given benchmark to train and test each model we studied. Given the relatively small size of our dataset, we chose this validation approach since this method is particularly advantageous for small datasets, as it maximizes the use of available data by allowing each data point to be used for both training and testing. It also reduces bias and variance by evaluating the model across multiple train–test splits and averaging the results, making the performance estimate less dependent on any single data partition. With a single holdout split, the estimate of performance depends heavily on how that one split happened to partition the data: if the test set is unrepresentative (for example, missing some rare information types), the estimate can be both biased and high-variance. In 10-fold cross-validation, each data point serves as test data exactly once and as training data nine times, so every region of the input space contributes to both training and evaluation, which tends to lower bias. Averaging the performance over 10 folds also smooths out randomness from any one unlucky split, which reduces variance in the estimate. Empirically, using around 10 folds is often a good trade-off: fewer folds (like 2–3) can give higher bias because the training sets are small, while very many folds (like leave-one-out) can increase variance because each test set is extremely small.\\


\subsubsection{Metrics Used}\label{sec:metrics}
We measured three metrics to report the performance of each model, Precision, Recall, and F1-score.
Precision ($P$) is the ratio of correctly predicted examples to all predicted examples for a given class.
Recall ($R$) is  the ratio of the number of correctly predicted examples and  all  examples of a given class. F1-score ($F1$) is the harmonic mean of
precision and recall.

{\scriptsize
\begin{eqnarray*}
P  = \frac{TP}{TP+FP},~
R = \frac{TP}{TP+FN},~
F1 = 2x\frac{P*R}{P+R}
\end{eqnarray*}}

TP = Correctly classified as an information type, 
FP = Incorrectly classified as a type, TN = Correctly classified as not a type, 
FN = Incorrectly classified as not an information type. 

We report these metrics using three averaging strategies: \textbf{macro average}, which computes the metric for each class independently and takes the unweighted mean, treating all information types equally regardless of their frequency; \textbf{micro average}, which computes metrics globally across all classes and is equivalent to overall accuracy in multi-class classification and \textbf{weighted average}, which weights each class by its support (number of instances), reflecting overall model performance on the dataset.

\subsection{Performance Analysis of Stand-alone Models (RQ1)} \label{sec:rq1}
\subsubsection{Motivation.} 
As part of our study on shallow ML models, we first adopted the best-performing model identified by Arya et al. \cite{InfoTypesOSSDiscussion} as our baseline classifier. Additionally, we explored another more advanced shallow ML model, LightGBM, to further enhance our ensemble classifier. We chose this model to take advantage of its highly efficient gradient-boosting decision tree architecture. It handles sparse data naturally and is faster with lower memory usage compared to other gradient boosting models. Given its high performance on structured data and its ability to handle feature interactions effectively, LightGBM was chosen for its efficiency in training and predicting, which is particularly beneficial in scenarios with a limited amount of data \cite{lightgbm}. We also focused on utilizing pre-trained transformer models (PTM) such as BERT \cite{bert} and RoBERTa \cite{roberta} to build deep ML classifiers, primarily due to their remarkable performance in natural language processing (NLP) tasks. Unlike shallow ML models, which often struggle to capture complex linguistic patterns and dependencies in text data, transformer models excel at understanding contextual information and capturing semantic nuances, thanks to their attention mechanism and deep architecture \cite{attentionisallyouneed}.

\subsubsection{Approach.} \label{sec:approach}
Here, we detail the development and optimization of the individual ML models—LightGBM, BERT, RoBERTa—and the baseline model for classifying information types in issue reports. BERT and RoBERTa perform deterministically with  fixed seeds (no sampling) while Random Forest and LightGBM are nondeterministic due to bagging/subsampling. All models are trained on individual sentences, with each sentence treated as a separate example.
\\ 
\bf{Random Forest (Baseline).} Arya et al.\cite{InfoTypesOSSDiscussion} explored a variety of configurations and feature sets to optimize the classification of information types in issue discussions. They implemented different balancing techniques, such as adjusting class weights or employing SMOTE (Synthetic Minority Over-Sampling Technique) to address class imbalances, alongside experimenting with textual and conversational features to comprehensively capture the semantics and contextual metadata of discussions. They primarily utilized Random Forest and Logistic Regression, which are commonly favored for text classification tasks due to their robustness and effectiveness. However, they excluded other classifiers like Support Vector Machines and Naive Bayes after preliminary testing indicated inferior performance and longer training times. Based on their findings, we adopted their best-performing model setup—Random Forest using class weight adjustments and conversational features—as our baseline. For this classifier, they considered  following hyperparameters in their optimization process: \begin{inparaenum}
\item {min\_samples\_split:} The minimum number of data points required to split a node. The values explored included 2, 5, and 10, aiming to determine the best threshold for creating new branches in the decision trees.
\item {n\_estimators:} The number of trees in the forest. By testing different quantities, such as 10, 50, and 100, they aimed to balance the model's complexity with its ability to generalize across different types of data.\end{inparaenum}

\bf {LightGBM (Light Gradient-Boosting Machine).} For this model, we represent the sentence text using TF-IDF features (top 10,000 terms) and concatenate them with conversational features to form the final input to the model. TF-IDF is chosen because it performs well on small, high-dimensional text datasets and captures sparse lexical patterns that complement the semantic information learned by BERT/RoBERTa in the ensemble.

To optimize the LightGBM model, we used Optuna~\cite{optuna}, an automatic hyperparameter tuning tool that employs Bayesian optimization to efficiently determine the best model parameters. Optuna was tasked with tuning the parameters listed in Table~\ref{tab:lgbm_params}, while we fixed parameters such as \textit{objective} to \texttt{multiclass} and \textit{metric} to \texttt{multi\_logloss} for this multi-class classification problem. Additionally, we addressed class imbalance by setting the \textit{class\_weight} parameter to \texttt{balanced}, ensuring that the model pays more attention to underrepresented classes. We also set \texttt{gbdt} (standard Gradient Boosting Decision Tree) as the \textit{boosting\_type}.

\begin{figure}[!t]
\centering
\includegraphics[scale=0.38]{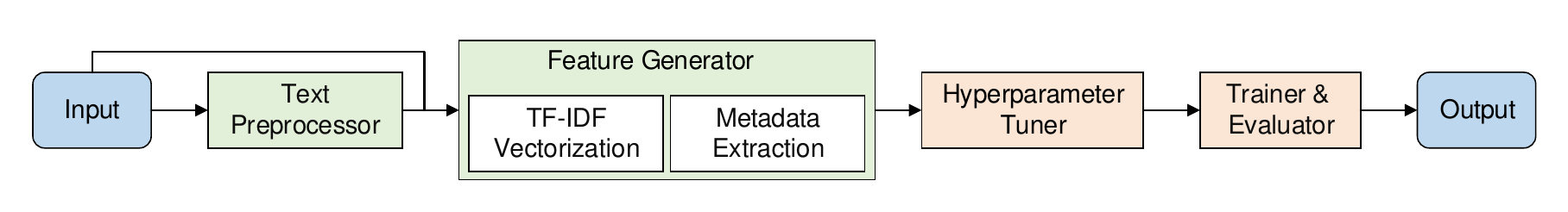}
\caption{LightGBM training architecture}
\label{fig:lightgbm}
\end{figure}

In addition to the above parameters, several other hyperparameters are also important in our setting because we train on relatively small, high‑dimensional, and imbalanced text datasets. The feature\_fraction and bagging\_fraction parameters regularize the model when using up to 10,000 TF‑IDF features, reducing overfitting and improving training speed by subsampling features and instances at each iteration. Furthermore, num\_leaves, min\_child\_samples, and lambda\_l1/l2 jointly control tree complexity and weight magnitudes, which is crucial to avoid overfitting given the limited number of labeled sentences in both datasets. We also enable feature\_pre\_filter to discard unused features early and tune bagging\_fraction for stochastic subsampling, further preventing overfitting. The model training architecture is shown in Figure \ref{fig:lightgbm}.

\begin{table}[!t]
\centering
\caption{HyperParameters for LightGBM classifiers}
\begin{tabular}{l p{65mm} l}
\toprule
\textbf{HyperParameter} & \textbf{Definition} & \textbf{Setting} \\
\midrule
num\_classes & Defines the number of target classes &  \\
objective & Specifies the learning task and the corresponding learning objective & multiclass \\
metric & Performance assessment metric to be minimized or maximized during training & multi\_logloss \\
boosting\_type & The type of algorithm to use for boosting & gbdt \\
class\_weight & Adjusts the weight of each class contribution to the objective & balanced \\
\midrule
feature\_pre\_filter & Whether to pre-filter features before building the dataset &  \\
lambda\_l1 & L1 regularization : effective in feature selection by shrinking some coefficients to zero &  \\
lambda\_l2 & L2 regularization : effective at combating the model's overfitting &  \\
feature\_fraction & The fraction of features to be used at each iteration &  \\
bagging\_fraction & The fraction of the data to be used for each iteration &  \\
bagging\_freq & Frequency for performing bagging at every N iterations &  \\
min\_child\_samples & Minimum number of data points needed in a leaf &  \\
num\_leaves & The maximum number of leaves in one tree &  \\
\bottomrule
\end{tabular}
\label{tab:lgbm_params}
\end{table}

\bf {BERT (Bidirectional Encoder Representations from Transformers).} BERT \cite{bert} is a powerful pre-trained model that uses transformer architecture to process words in relation to all other words in a sentence rather than one-by-one in order. This allows BERT to understand the context of a word based on all of its surroundings, which is a significant advantage for classifying information types in textual data where the context in which words appear is crucial. The bidirectional nature of BERT makes it adept at understanding the nuances and complexities of language in issue reports. For instance, distinguishing between \textit{Motivation} and \textit{Expected Behaviour} often requires examining the context following the key phrase. Consider the examples: 
``I expect it to fail because the input is invalid,'' which corresponds to \textit{Expected Behaviour}, and 
``We should support this use case because users often need …,'' which reflects \textit{Motivation}. 
In these cases, phrases such as ``because the input is invalid'' and ``because users often need'' provide crucial contextual cues that appear later in the sentence. A bidirectional encoder can attend to both the preceding and following context simultaneously, enabling it to better disambiguate between these labels.

\bf{RoBERTa (Robustly Optimized BERT Approach).} RoBERTa \cite{roberta} builds on BERT by modifying key hyperparameters, removing the next-sentence pretraining objective and training with much larger mini-batches and learning rates. This results in an improved model that outperforms BERT on many benchmarking tests. We chose RoBERTa because it offers all the benefits of BERT but with refinements that increase its accuracy and robustness in understanding textual data. The comparison between architecture details of BERT and RoBERTa Model is presented in \tbl\ref{tab:archbertroberta}.

\begin{table}[!t]
    \centering
    \caption{Architecture details of BERT and RoBERTa model (L = Layers, H = Hidden, D = Heads, P = Parameters)}
    \label{tab:archbertroberta}
    \begin{tabular}{llcccc}
        \toprule
        \textbf{Architecture} & \textbf{Pre-trained Checkpoint} & \textbf{L} & \textbf{H} & \textbf{D} & \textbf{P} \\
        \midrule
        BERT & bert-base-uncased & 12 & 768 & 12 & 110M \\
        RoBERTa & roberta-base & 12 & 768 & 12 & 125M \\
        \bottomrule
    \end{tabular}
\end{table}

\begin{table}[!t]
\caption{Stand-alone models and inputs}
    \centering
    \begin{tabular}{ll}
     \toprule
    \textbf{Model} & \textbf{Input}\\ 
     \midrule 
     Random Forest (Baseline) & Conversational Features \\
     LightGBM & Textual + Conversational Features \\   
     BERT  & Preprocessed Text \\ 
     ROBERTA & Preprocessed Text  \\
     \bottomrule
    \end{tabular}
    \label{tab:model_table}
\end{table}

\begin{figure}[!t]
\centering
\includegraphics[scale=0.38]{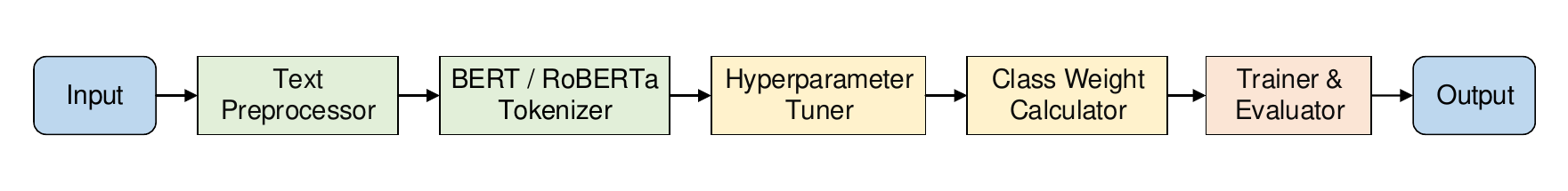}
\caption{BERT/RoBERTa fine-tuning  architecture}
\label{fig:dlclassifier}
\end{figure}

As input to our deep ML models, we utilized preprocessed text (see Table \ref{tab:model_table}), which was tokenized using either the \it{bert-base-uncased} or \it{roberta-based} tokenizer, depending on the selected model variant. The text preprocessing details are provided in Section \ref{sec:featureengr}. Additionally, we generated a validation set by extracting 5\% of the training data. To address class imbalances, we computed class weights based on the frequency of each label category within the training data as
\(\text{Weight}_c = \frac{N}{K \cdot n_c}\), 
where \(N\) is the total number of training examples, \(K\) is the number of classes, and \(n_c\) is the number of examples in class \(c\). This corresponds to class\_weight = balanced in our shallow classifiers. These weights were then incorporated into the model training process to mitigate bias towards the majority class. Our model training pipeline utilized a custom trainer class, which computed the loss function while considering the assigned class weights during the forward pass. The hyperparameters for fine-tuning were selected based on recommendations from the original BERT~\cite{bert} and RoBERTa~\cite{roberta} papers and further optimized via grid search over learning rate \{1e-5, 2e-5, 3e-5\} and batch size \{8, 16\}. This process yielded a final configuration of learning rate = 2e-5 and batch size = 16. The final layer is a dense linear classification head on the [CLS] embedding, producing a dense logit vector with one value per class (13 for Dataset 1, 12 for Dataset 2). Softmax converts the logits into class probabilities, and the predicted label is the argmax. The training architectures of these models are shown in \fig\ref{fig:dlclassifier}.

\subsubsection{Results} \label{sec:results}
In Tables \ref{tab:result_individual_d1} and \ref{tab:result_individual_d2}, we detail the performance metrics—precision, recall, and F1-score—for each information type and overall, comparing stand-alone ML models across two datasets. Tables \ref{tab:fold_result_individual_d1} and \ref{tab:fold_result_individual_d2} report the performance of the same models on each fold of datasets 1 and 2. Notably, LightGBM surpasses the baseline model (Random Forest) \cite{InfoTypesOSSDiscussion}, achieving the highest F1 score in Dataset 1 at 0.67, showcasing the strength of gradient boosting techniques in processing feature-dense textual data. This surpasses the baseline's F1 score of 0.61. BERT and RoBERTa, however, scored lower with F1 scores of 0.54 and 0.56, respectively. In Dataset 2, all models improved upon the baseline, with scores ranging between 0.61-0.63, suggesting that dataset 2 benefits more from text embeddings, with LightGBM again leading. This variation in performance between the datasets suggests that they differ significantly in content and structure. Dataset 2, unlike Dataset 1, appears to benefit more from text embeddings, likely due to its content having more specialized vocabulary or unique linguistic patterns. These embeddings can grasp the subtle meanings of words within the specific context of the discussions, providing an advantage in understanding and classifying the information types more accurately.

\begin{table}[!t]
    \centering
    \caption{Performance of stand-alone ML models on dataset 1.}
    \label{tab:result_individual_d1}
    \begin{adjustbox}{width=\columnwidth}
    \centering
    \fontsize{6pt}{6pt}\selectfont
    \begin{tabular}{p{2cm}|p{.22cm}p{.22cm}c|p{.22cm}p{.22cm}c|p{.22cm}p{.22cm}c|p{.22cm}p{.22cm}c|p{.25cm}}
    \toprule
        \bf{Dataset 1} & \multicolumn{3}{c|}{\textbf{Baseline (RF)}} & \multicolumn{3}{c|}{\textbf{LightGBM}} & \multicolumn{3}{c|}{\textbf{BERT}} & \multicolumn{3}{c|}{\textbf{RoBERTa}} \\
        Information Type & Pr & Rc & F1 & Pr & Rc & F1 & Pr & Rc & F1  & Pr & Rc & F1  & Sup \\
        \midrule
        Action on Issue                 & 0.45 & 0.39 & 0.42    & 0.9 & 0.46 & 0.6     & 0.76 & 0.67 & \bf{0.69} & 0.7 & 0.7 & \bf{0.69} & 61 \\
        Bug Reproduction                & 0.53 & 0.47 & 0.5    & 0.68 & 0.62 & \bf{0.65}   & 0.51 & 0.5 & 0.5   & 0.49 & 0.59 & 0.53    & 245 \\
        Contribution \& Commit. & 0.25 & 0.19 & 0.21 & 0.56 & 0.19 & 0.26 & 0.58 & 0.48 & \bf{0.51} & 0.53 & 0.52 & \bf{0.51} & 83 \\
        Expected Behaviour & 0.42 & 0.28 & 0.33 &   0.59 & 0.41 & \bf{0.46}  & 0.34 & 0.32 & 0.33 & 0.42 & 0.42 & 0.42 & 124 \\
        Investigation and Explr.  & 0.6 & 0.65 & 0.62 &    0.68 & 0.65 & \bf{0.66}  & 0.43 & 0.47 & 0.45 & 0.46 & 0.5 & 0.48 & 377 \\
        Motivation & 0.56 & 0.53 & \bf{0.54} &   0.63 & 0.47 & 0.53 &    0.31 & 0.36 & 0.34 &    0.38 & 0.41 & 0.39  & 288 \\
        Observed Bug Behav. & 0.56 & 0.7 & 0.62 &     0.75 & 0.69 & \bf{0.69} &   0.4 & 0.27 & 0.32 &     0.33 & 0.28 & 0.3 & 131 \\
        Potential New Issues. & 0.67 & 0.66 & 0.66 &    0.74 & 0.68 & \bf{0.7} &     0.23 & 0.23 & 0.23 &    0.26 & 0.21 & 0.23 & 230 \\
        Social Conversation & 0.63 & 0.62 & 0.63 & 0.71 & 0.71 & 0.71 & 0.82 & 0.78 & 0.8 & 0.84 & 0.81 & \bf{0.83} & 798 \\
        Solution Discussion & 0.68 & 0.71 & 0.69 &  0.67 & 0.82 & \bf{0.74} &    0.61 & 0.58 & 0.59 &    0.63 & 0.55 & 0.59 & 1411 \\
        Task Progress & 0.27 & 0.14 & 0.18 &    0.5 & 0.16 & 0.23   & 0.47 & 0.43 & 0.44    & 0.5 & 0.5 & \bf{0.48} & 125 \\
        Usage & 0.65 & 0.82 & 0.73 & 0.75 & 0.8 & \bf{0.77} & 0.49 & 0.55 & 0.51 & 0.46 & 0.6 & 0.52 & 368 \\
        Workarounds & 0.58 & 0.45 & 0.49 & 0.84 & 0.44 & \bf{0.56} & 0.31 & 0.31 & 0.31 & 0.33 & 0.38 & 0.35 & 89 \\
        \midrule
        \bf{Micro Average} & 0.64 & 0.64 & 0.64 & 0.68 & 0.68 & 0.68 & 0.54 & 0.54 & 0.54 & 0.56 & 0.56 & 0.56 & 4330 \\
        \bf{Macro Average} & 0.56 & 0.54 & 0.54 & 0.69 & 0.54 & 0.58 & 0.48 & 0.46 & 0.46 & 0.49 & 0.5 & 0.48 & 4330 \\
        \bf{Weighted Average} & 0.61 & 0.62 & 0.61 & 0.68 & 0.68 & \bf{0.67} & 0.55 & 0.54 & 0.54 & 0.57 & 0.56 & 0.56 & 4330 \\
        \bottomrule
    \end{tabular}
    \end{adjustbox}
\end{table}

\begin{table}[!t]
    \centering
    \caption{Performance of stand-alone ML models on dataset 2}
    \label{tab:result_individual_d2}
        \begin{adjustbox}{width=\columnwidth}
    \centering
    \fontsize{6pt}{6pt}\selectfont
    \begin{tabular}{p{2cm}|p{.22cm}p{.22cm}c|p{.22cm}p{.22cm}c|p{.22cm}p{.22cm}c|p{.22cm}p{.22cm}c|p{.25cm}}
        \toprule
        \bf{Dataset 2} & \multicolumn{3}{c|}{\textbf{Baseline (RF)}} & \multicolumn{3}{c|}{\textbf{LightGBM}} & \multicolumn{3}{c|}{\textbf{BERT}} & \multicolumn{3}{c|}{\textbf{RoBERTa}} \\
        Information Type & Pr & Rc & F1  & Pr & Rc & F1  & Pr & Rc & F1  & Pr & Rc & F1  & Sup \\
        \midrule
        Action on Issue & 0.28 & 0.2 & 0.23 & 0.66 & 0.44 & 0.51 & 0.7 & 0.75 & \textbf{0.72} & 0.66 & 0.79 & \textbf{0.72} & 96 \\
        Bug Reproduction & 0.4 & 0.18 & 0.24 & 0.65 & 0.4 & 0.49 & 0.58 & 0.53 & 0.54 & 0.56 & 0.57 & \textbf{0.57} & 176 \\
        Contribution \& Commit. & 0.23 & 0.11 & 0.15 & 0.73 & 0.36 & 0.47 & 0.68 & 0.59 & 0.6 & 0.67 & 0.65 & \textbf{0.64} & 109 \\
        Expected Behaviour & 0.33 & 0.21 & 0.25 & 0.12 & 0.08 & 0.1 & 0.24 & 0.17 & 0.19 & 0.43 & 0.33 & \textbf{0.32} & 24 \\
        Investigation and Explr. & 0.63 & 0.83 & 0.71 & 0.65 & 0.89 & \textbf{0.75} & 0.66 & 0.68 & 0.67 & 0.69 & 0.67 & 0.68 & 1891 \\
        Motivation & 0.56 & 0.44 & 0.46 & 0.83 & 0.56 & \textbf{0.65} & 0.28 & 0.22 & 0.24 & 0.17 & 0.11 & 0.13 & 18 \\
        Observed Bug Behav & 0.37 & 0.3 & 0.31 & 0.48 & 0.27 & \textbf{0.34 }& 0.23 & 0.23 & 0.23 & 0.28 & 0.31 & 0.29 & 81 \\
        Social Conversation & 0.43 & 0.37 & 0.4 & 0.62 & 0.44 & 0.52 & 0.66 & 0.56 & 0.6 & 0.68 & 0.58 & \textbf{0.62} & 469 \\
        Solution Discussion & 0.62 & 0.58 & 0.6 & 0.7 & 0.6 & \textbf{0.64} & 0.54 & 0.58 & 0.56 & 0.54 & 0.6 & 0.57 & 856 \\
        Task Progress & 0.33 & 0.16 & 0.21 & 0.43 & 0.16 & 0.23 & 0.44 & 0.43 & 0.43 & 0.46 & 0.48 & \textbf{0.47} & 233 \\
        Testing & 0.57 & 0.31 & 0.39 & 0.68 & 0.51 & 0.58 & 0.65 & 0.58 & 0.61 & 0.64 & 0.58 & \textbf{0.6} & 261 \\
        Workarounds & 0.31 & 0.18 & 0.23 & 0.72 & 0.59 & 0.62 & 0.64 & 0.59 & 0.6 & 0.71 & 0.7 & \textbf{0.67} & 27 \\
        \midrule
        \bf{Micro Average} & 0.58 & 0.58 & 0.58 & 0.65 & 0.65 & 0.65 & 0.61 & 0.61 & 0.61 & 0.62 & 0.62 & 0.62 & 4241 \\
        \bf{Macro Average} & 0.42 & 0.32 & 0.34 & 0.61 & 0.44 & 0.49 & 0.52 & 0.49 & 0.5 & 0.53 & 0.52 & 0.52 & 4241 \\
        \textbf{Weighted Average} & 0.55 & 0.58 & 0.55 & 0.64 & 0.65 & \textbf{0.63} & 0.61 & 0.61 & 0.6 & 0.63 & 0.62 & 0.62 & 4241 \\
        \bottomrule
    \end{tabular}
    \end{adjustbox}
\end{table}

The F1 scores for LightGBM, BERT, and RoBERTa across various labels provide insightful reflections on each model's handling of different types of information in issue discussions. BERT and RoBERTa, known for their deep learning capabilities and handling of natural language through contextual embeddings, show better performance in Action on Issue, Contribution and Commitment, Social Conversation, and Task Progress in both datasets. These categories benefit from the models' ability to understand natural language. However, there are some cases where text-based models get confused due to the vocabulary ambiguity. \textit{'This issue should not be labeled as a stop-ship/blocker issue for 0.29.1 due to the presence of a workaround'}; this sentence gets labeled as Workarounds though it should be of Action on Issue. Because of the word \it{workaround}, the text-based models got confused. On the other hand, shallow ML models typically achieve better performance on Investigation and Exploration, Motivation, Observed Bug Behaviour and Solution Discussion in both datasets, indicating their potential in leveraging conversational features like code snippets, the length of the discussion, the position of sentences within the conversation and others. For example, while training, Observed Bug Behaviour are mostly found on the start of the conversation. These models learned to give more importance to the positions of the sentences in the thread when labeling this class. Also, Investigation and Exploration, Solution Discussion and Testing give importance to the existence of code. Again, Solution Discussion related sentences are much longer than Social conversation and Contribution and Commitment related sentences. Investigation and Exploration comments generally come from the authors who are contributors to the repository. However, these models struggle with Contribution and Commitment and Task Progress, where they may not have sufficient distinctive textual or conversational cues to differentiate between categories effectively. This also indicates a limitation in the model's ability to capture the more subtle nuances of language that could be critical for these information types.

\begin{table}[!t]
    \centering
    \caption{Performance of stand-alone ML models on each fold of dataset 1.}
    \label{tab:fold_result_individual_d1}
    \begin{adjustbox}{width=\columnwidth}
    \centering
    \fontsize{6pt}{6pt}\selectfont
    \begin{tabular}{c|p{.4cm}p{.4cm}c|p{.4cm}p{.4cm}c|p{.4cm}p{.4cm}c|p{.4cm}p{.4cm}c}
    \toprule
 & \multicolumn{3}{c|}{\textbf{Baseline (RF)}} & \multicolumn{3}{c|}{\textbf{LightGBM}} & \multicolumn{3}{c|}{\textbf{BERT}} & \multicolumn{3}{c}{\textbf{RoBERTa}} \\ 
        
Fold & Pr & Rc & F1 & Pr & Rc & F1 & Pr & Rc & F1 & Pr & Rc & F1 \\
\hline
1  & 0.60 & 0.62 & 0.61 & 0.69 & 0.69 & 0.68 & 0.53 & 0.53 & 0.53 & 0.56 & 0.56 & 0.55 \\
2  & 0.63 & 0.63 & 0.62 & 0.68 & 0.69 & 0.67 & 0.52 & 0.50 & 0.50 & 0.55 & 0.55 & 0.55 \\
3  & 0.64 & 0.65 & 0.64 & 0.67 & 0.67 & 0.66 & 0.55 & 0.53 & 0.54 & 0.59 & 0.57 & 0.57 \\
4  & 0.64 & 0.64 & 0.63 & 0.68 & 0.68 & 0.67 & 0.55 & 0.55 & 0.55 & 0.60 & 0.55 & 0.56 \\
5  & 0.63 & 0.65 & 0.64 & 0.68 & 0.68 & 0.67 & 0.57 & 0.55 & 0.55 & 0.55 & 0.54 & 0.54 \\
6  & 0.62 & 0.62 & 0.62 & 0.67 & 0.66 & 0.65 & 0.54 & 0.55 & 0.54 & 0.55 & 0.55 & 0.55 \\
7  & 0.66 & 0.67 & 0.66 & 0.76 & 0.73 & 0.72 & 0.55 & 0.54 & 0.54 & 0.57 & 0.56 & 0.56 \\
8  & 0.60 & 0.61 & 0.60 & 0.64 & 0.64 & 0.63 & 0.55 & 0.53 & 0.53 & 0.58 & 0.57 & 0.57 \\
9  & 0.65 & 0.64 & 0.64 & 0.69 & 0.70 & 0.69 & 0.59 & 0.58 & 0.58 & 0.59 & 0.58 & 0.58 \\
10 & 0.64 & 0.64 & 0.64 & 0.68 & 0.67 & 0.66 & 0.57 & 0.58 & 0.57 & 0.57 & 0.57 & 0.57 \\
\midrule
Mean     & 0.63 & 0.64 & 0.63 & 0.68 & 0.68 & 0.67 & 0.55 & 0.54 & 0.54 & 0.57 & 0.56 & 0.56 \\
Variance & 0.0003 & 0.0003 & 0.0003 & 0.0008 & 0.0005 & 0.0005 & 0.0004 & 0.0006 & 0.0004 & 0.0003 & 0.0002 & 0.0001 \\
\bottomrule
\end{tabular}
\end{adjustbox}
\end{table}
\begin{table}[!t]
    \centering
    \caption{Performance of stand-alone ML models on each fold of dataset 2.}
    \label{tab:fold_result_individual_d2}
    \begin{adjustbox}{width=\columnwidth}
    \centering
    \fontsize{6pt}{6pt}\selectfont
    \begin{tabular}{c|p{.4cm}p{.4cm}c|p{.4cm}p{.4cm}c|p{.4cm}p{.4cm}c|p{.4cm}p{.4cm}c}
    \toprule
 & \multicolumn{3}{c|}{\textbf{Baseline (RF)}} & \multicolumn{3}{c|}{\textbf{LightGBM}} & \multicolumn{3}{c|}{\textbf{BERT}} & \multicolumn{3}{c}{\textbf{RoBERTa}} \\ 
        
Fold & Pr & Rc & F1 & Pr & Rc & F1 & Pr & Rc & F1 & Pr & Rc & F1 \\
\hline
1  & 0.54 & 0.57 & 0.54 & 0.64 & 0.64 & 0.62 & 0.61 & 0.61 & 0.60 & 0.60 & 0.60 & 0.59 \\
2  & 0.51 & 0.55 & 0.52 & 0.65 & 0.67 & 0.64 & 0.60 & 0.59 & 0.59 & 0.60 & 0.60 & 0.60 \\
3  & 0.56 & 0.58 & 0.56 & 0.61 & 0.63 & 0.60 & 0.56 & 0.54 & 0.54 & 0.58 & 0.57 & 0.57 \\
4  & 0.56 & 0.57 & 0.55 & 0.68 & 0.68 & 0.66 & 0.63 & 0.64 & 0.63 & 0.67 & 0.66 & 0.66 \\
5  & 0.53 & 0.59 & 0.55 & 0.62 & 0.65 & 0.62 & 0.58 & 0.58 & 0.58 & 0.62 & 0.61 & 0.61 \\
6  & 0.57 & 0.59 & 0.57 & 0.69 & 0.68 & 0.66 & 0.64 & 0.64 & 0.64 & 0.64 & 0.63 & 0.63 \\
7  & 0.54 & 0.58 & 0.54 & 0.62 & 0.64 & 0.61 & 0.61 & 0.60 & 0.61 & 0.66 & 0.65 & 0.65 \\
8  & 0.58 & 0.59 & 0.56 & 0.65 & 0.65 & 0.63 & 0.64 & 0.64 & 0.64 & 0.66 & 0.65 & 0.65 \\
9  & 0.55 & 0.59 & 0.56 & 0.65 & 0.66 & 0.64 & 0.59 & 0.59 & 0.59 & 0.65 & 0.64 & 0.64 \\
10 & 0.54 & 0.57 & 0.54 & 0.63 & 0.64 & 0.61 & 0.62 & 0.61 & 0.61 & 0.59 & 0.59 & 0.59 \\
\midrule
Mean     & 0.548 & 0.578 & 0.549 & 0.644 & 0.654 & 0.629 & 0.61 & 0.61 & 0.60 & 0.63 & 0.62 & 0.62 \\
Variance & 0.0004 & 0.0002 & 0.0002 & 0.0006 & 0.0003 & 0.0004 & 0.0007 & 0.0009 & 0.0008 & 0.0009 & 0.0008 & 0.0009 \\
\bottomrule
\end{tabular}
\end{adjustbox}
\end{table}

In all cases, LightGBM outperforms Random Forest due to its efficient gradient boosting method, which corrects previous trees' errors, handles sparse data effectively and uses textual features. Again, RoBERTa shows better performance than BERT due to its improved pre-training regime, including training with larger mini-batches and removing the next-sentence prediction task, which enables it to capture linguistic nuances more effectively. Our fine-tuning setup for BERT and RoBERTa remains largely comparable, using the same weighted cross-entropy objective with only minor differences in learning rate.

We further examine recall to understand how well each model identifies critical information types. On Dataset 1, LightGBM achieves strong recall for \textit{Solution Discussion} (0.82) and \textit{Observed Bug Behaviour} (0.69), demonstrating effective detection of these essential categories. \textit{Bug Reproduction} recall is moderate at 0.62, and \textit{Workarounds} recall at 0.44 reflects the challenge of detecting this relatively rare category (support=89). \textit{Task Progress} recall varies across models (0.14--0.50), consistent with its small support and subtle linguistic cues. On Dataset 2, recall for several types decreases slightly, with \textit{Bug Reproduction} at 0.40 for LightGBM, \textit{Observed Bug Behaviour} at 0.27, and \textit{Motivation} at 0.56, reflecting the added challenge of the specialized OpenJ9 vocabulary - a pattern consistent across all models on this dataset.


\subsubsection{Misclassification Analysis}
\begin{table}[!t]
    \centering
    \caption{How the misclassification of a classifier can be corrected by one or at least one classifier}
    \label{tab:wrong_stand-alone}
    \begin{tabular}{l|cccc|c}
    \toprule
        \textbf{Stand-alone Models} & \textbf{RF} & \textbf{LGBM} & \textbf{BERT} & \textbf{RoBERTa} & \textbf{$\geq 1$} \\
        \midrule
         \multicolumn{6}{c}{\textbf{Dataset 1}}\\
        \midrule
        Random Forest & ~ & 34\% & 47\% & 49\% & 68\% \\ 
        LightGBM & 25\% & ~ & 36\% & 39\% & 64\% \\ 
        BERT & 58\% & 56\% & ~ & 28\% & 75\% \\ 
        RoBERTa & 58\% & 56\% & 26\% & ~ & 74\% \\ 
        \midrule
        \multicolumn{6}{c}{\textbf{Dataset 2}}\\
        \midrule
        Random Forest & ~ & 35\% & 50\% & 54\% & 69\% \\ 
        LightGBM & 21\% & ~ & 39\% & 43\% & 62\% \\ 
        BERT & 47\% & 46\% & ~ & 30\% & 66\% \\ 
        RoBERTa & 49\% & 48\% & 28\% & ~ & 65\% \\ 
        \bottomrule
    \end{tabular}
\end{table}

 Table \ref{tab:wrong_stand-alone} presents how misclassifications by one stand-alone model can be corrected by other models, with the last column ($\geq 1$) indicating how the misclassification of a model can be corrected by the correct classification of at least one of the models. This analysis underlines the potential benefits of using an ensemble approach where the weaknesses of one model are compensated by the strengths of others. For dataset 1, the data shows that the misclassifications made by BERT and RoBERTa are more frequently corrected by other models, with both achieving correction rates of over 70\% when at least one other model is considered. This suggests that while BERT and RoBERTa might be prone to specific errors, their weaknesses are effectively covered by the capabilities of the other models, particularly Random Forest and LightGBM. Random Forest and LightGBM themselves show a moderate level of error correction by others, indicating a more balanced level of dependency across models. For dataset 2, the correction rates are generally lower, which might be indicative of more challenging or ambiguous data that even complementary models struggle to correct. However, for both datasets, all models still benefit from corrections by others, with rates over 60\%. This reflects the ongoing relevance of an ensemble approach, especially in complex datasets where no single model is perfect. Overall, these results justify the use of an ensemble approach in scenarios where diverse models can leverage their unique strengths, leading to improved overall performance by filling in each other's gaps. 

 \begin{tcolorbox}[flushleft upper, boxrule=1pt, arc=0pt, left=0pt, right=0pt, top=0pt, bottom=0pt, colback=white, after=\ignorespacesafterend\par\noindent]
\textbf{Summary of RQ1.} 
Among the stand-alone models, in both datasets, LightGBM achieves the highest F1-score for information type detection in both datasets, surpassing SOTA model~\cite{InfoTypesOSSDiscussion} by 6-8\%. 
\end{tcolorbox}
\subsection{Performance Analysis of Ensemble Models (RQ2)} \label{sec:rq2} 
\subsubsection{Motivation.} \label{sec:rq2motivation}  To further enhance the performance of the overall issue discussion information type classification, we thought of leveraging distinct strengths and mitigating individual weaknesses of each stand-alone model. Also, \tbl\ref{tab:wrong_stand-alone} shows how the misclassification of a classifier can be corrected by one or at least one classifier. Therefore, we used these four models - Random Forest \cite{InfoTypesOSSDiscussion}, LightGBM, BERT, RoBERTa, to make an ensemble classifier for information type detection. This approach is suitable when the development environment can  deploy multiple models that form an ensemble with the hope of producing better performance than the stand-alone models.

The models are chosen because no single one sees all the signal in our data, and each fills a gap the others leave open. Random Forest and LightGBM work from explicit, engineered features and combine trees differently (bagging vs. boosting), giving each a different bias-variance profile and access to different features: Random Forest is robust to noisy, heterogeneous, low-dimensional metadata without assuming linearity \cite{breiman2001random}, while LightGBM's histogram-based leaf-wise boosting scales to the high dimensional sparse TF-IDF features and captures non-linear feature interactions, with built-in L1/L2 regularization (Table \ref{tab:lgbm_params}) controlling overfitting on our limited labeled data \cite{lightgbm, friedman2001greedy}. Gradient-boosted trees also remain competitive with deep models on small, high-dimensional sparse problems like ours \cite{grinsztajn2022why}.

BERT \cite{bert} and RoBERTa \cite{roberta}  see the raw sentence text directly and learn context-dependent meaning through self-attention, rather than relying on hand-engineered features. This is what lets them catch cases the tree models miss: two sentences with identical surface vocabulary but different meaning depending on context (e.g., our Motivation vs. Expected Behaviour example). Conversely, the tree models catch what the transformers miss, namely conversational and structural cues like turn position or author role that never enter the transformers' input at all.

We include both BERT and RoBERTa because RoBERTa differs from BERT only in pretraining regime, not architecture \cite{roberta}, which is enough on its own to produce decorrelated errors, since ensembling benefits from base learners disagreeing rather than from architectural novelty \cite{wolpert1992stacked, dietterich2000ensemble, kuncheva2003measures}. Our per-class results (Tables ~\ref{tab:result_individual_d1} -~\ref{tab:result_individual_d2}) confirm this empirically: BERT and RoBERTa are strongest on different information types, and no model dominates across all types, which is the practical evidence that the ensemble's complementary design pays off.

\subsubsection{Approach.} \label{sec:rq2approach} 
\begin{figure}[!t]
\centering
\includegraphics[scale=0.6]{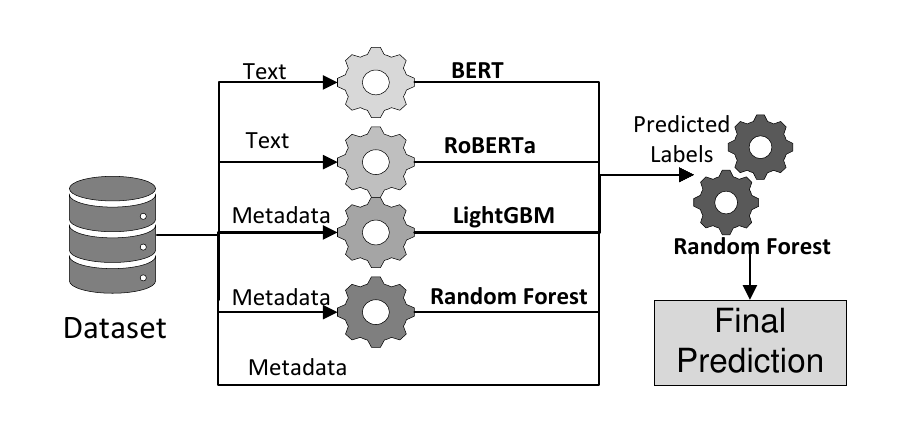}
\caption{The architecture of ensemble classifier}
\label{fig:iitd}
\end{figure}

To construct the ensemble classifier, the predicted labels from four distinct classifiers are combined with textual and conversational features from the dataset to feed into a ML classifier. This classifier then generates the final prediction. We use hard predicted labels from stand-alone models rather than their predicted probabilities, to keep the ensemble simple and reduce overfitting risks on our $\sim$4k sentences. Passing soft probabilities can propagate correlated errors from similar base models and amplify rather than diversify their predictions. We perform one 10-fold cross-validation per dataset. There is no separate, second cross-validation for the ensemble. In each fold, base models train on the other nine folds. They predict only on the held-out fold. So their predictions feeding the ensemble are always out-of-fold. The ensemble is trained and evaluated on this same fold structure. No sentence's ensemble prediction relies on a base model trained on that sentence. This holds independently for each dataset. The two datasets are never mixed. The structure of this ensemble approach is illustrated in Figure \ref{fig:iitd}. 

For the final classification step, we have experimented with a variety of classifiers, including Random Forest, LightGBM, BERT, and RoBERTa. The technique of automatic hyperparameter tuning previously detailed in Section \ref{sec:rq1} is applied to the shallow ML models. In contrast, for deep learning models like BERT and RoBERTa, we enhance input integration by prefixing each data point with predicted labels from stand-alone models, formatted as \it{PRED\_ModelName\_Label}. These models are subsequently fine-tuned with carefully selected hyperparameters to optimize performance with batch size 8 and epoch 10. As learning rate, we chose 5e-5 and 3e-5 for BERT and RoBERTa respectively. 

\subsubsection{Results}\label{sec:ensembleResults}
Tables \ref{tab:result_ensemble_d1} and \ref{tab:result_ensemble_d2} demonstrate the efficacy of ensemble classifiers across two distinct datasets. Tables \ref{tab:fold_result_ensemble_d1} and \ref{tab:fold_result_ensemble_d2} report the performance of the same models on each fold of datasets 1 and 2. On both datasets, the best ensemble model surpasses the baseline models and all other stand-alone implementations and establish the capability of ensemble methods complementing any stand alone models.

\begin{table}[!t]
    \centering
    \caption{Performance of ensemble classifiers on dataset 1. (RF = Random Forest, LG = LightGBM, B = BERT, RB = RoBERTa)}
    \label{tab:result_ensemble_d1}
    \begin{adjustbox}{width=\columnwidth}
    \centering
    \fontsize{6pt}{6pt}\selectfont
    \begin{tabular}{p{2cm}|p{.22cm}p{.22cm}c|p{.22cm}p{.22cm}c|p{.22cm}p{.22cm}c|p{.22cm}p{.22cm}c|p{.25cm}}

    \toprule
         \bf{Dataset 1} & \multicolumn{3}{c|}{\textbf{Ensemble (RF)}} & \multicolumn{3}{c|}{\textbf{Ensemble (LG)}} & \multicolumn{3}{c|}{\textbf{Ensemble (B)}} & \multicolumn{3}{c|}{\textbf{Ensemble (RB)}}\\
        Information Type & Pr & Rc & F1  & Pr & Rc & F1  & Pr & Rc & F1  & Pr & Rc & F1  & Sup \\ 
        \midrule
        Action on Issue & 0.69 & 0.8 & 0.72 & 0.83 & 0.72 & \textbf{0.74} & 0.76 & 0.69 & 0.71 & 0.61 & 0.75 & 0.65 & 61 \\ 
        Bug Reproduction & 0.64 & 0.68 & 0.65 & 0.72 & 0.69 & \textbf{0.7} & 0.68 & 0.64 & 0.66 & 0.64 & 0.75 & 0.69 & 245 \\ 
        Contribution \& Commit. & 0.56 & 0.41 & 0.45 & 0.64 & 0.41 & 0.46 & 0.48 & 0.47 & 0.47 & 0.47 & 0.65 & \textbf{0.54} & 83 \\ 
        Expected Behaviour & 0.58 & 0.47 & 0.51 & 0.64 & 0.52 & \textbf{0.57} & 0.46 & 0.43 & 0.43 & 0.53 & 0.56 & 0.54 & 124 \\ 
        Investigation and Explr. & 0.67 & 0.67 & 0.67 & 0.7 & 0.72 &\textbf{ 0.7} & 0.68 & 0.66 & 0.67 & 0.63 & 0.69 & 0.66 & 377 \\ 
        Motivation & 0.6 & 0.61 & \textbf{0.6} & 0.66 & 0.56 & \textbf{0.6} & 0.56 & 0.58 & 0.57 & 0.56 & 0.61 & 0.58 & 288 \\ 
        Observed Bug Behav. & 0.69 & 0.73 & 0.7 & 0.77 & 0.71 & \textbf{0.72} & 0.76 & 0.67 & 0.69 & 0.75 & 0.7 & 0.71 & 131 \\ 
        Potential New Issues. & 0.73 & 0.69 & \textbf{0.7 }& 0.72 & 0.65 & 0.67 & 0.67 & 0.65 & 0.66 & 0.66 & 0.71 & 0.68 & 230 \\ 
        Social Conversation & 0.83 & 0.74 & 0.79 & 0.82 & 0.79 & 0.8 & 0.84 & 0.78 & \textbf{0.81} & 0.85 & 0.77 & 0.8 & 798 \\ 
        Solution Discussion & 0.75 & 0.79 & \textbf{0.77} & 0.73 & 0.81 & \textbf{0.77} & 0.74 & 0.77 & 0.76 & 0.82 & 0.7 & 0.75 & 1411 \\ 
        Task Progress & 0.61 & 0.29 & 0.38 & 0.52 & 0.34 & 0.4 & 0.47 & 0.43 & 0.45 & 0.49 & 0.46 & \textbf{0.46} & 125 \\ 
        Usage & 0.74 & 0.83 & \textbf{0.78} & 0.74 & 0.79 & 0.76 & 0.73 & 0.79 & 0.76 & 0.73 & 0.84 & \textbf{0.78} & 368 \\ 
        Workarounds & 0.68 & 0.6 & \textbf{0.62} & 0.67 & 0.52 & 0.58 & 0.61 & 0.56 & 0.58 & 0.59 & 0.69 & \textbf{0.62} & 89 \\ 
        \midrule
        \bf{Micro Average} & 0.72 & 0.72 & 0.72 & 0.73 & 0.73 & 0.73 & 0.71 & 0.71 & 0.71 & 0.71 & 0.71 & 0.71& 4330 \\ 
        \bf{Macro Average} & 0.67 & 0.64 & 0.64 & 0.71 & 0.63 & 0.65 & 0.65 & 0.62 & 0.63 & 0.64 & 0.68 & 0.65 & 4330 \\ 
        \bf{Weighted Average} & 0.72 & 0.72 & 0.71 & 0.73 & 0.72 & \bf{0.72} & 0.71 & 0.7 & 0.7 & 0.73 & 0.71 & 0.71 & 4330 \\ 
        \bottomrule
    \end{tabular}
    \end{adjustbox}
\end{table}
\begin{table}[!t]
    \centering
    \caption{Performance of ensemble classifiers on dataset 2. (RF = Random Forest, LG = LightGBM, B = BERT, RB = RoBERTa)}
    \label{tab:result_ensemble_d2}
    \begin{adjustbox}{width=\columnwidth}
    \centering
    \fontsize{6pt}{6pt}\selectfont
    \begin{tabular}{p{2cm}|p{.22cm}p{.22cm}c|p{.22cm}p{.22cm}c|p{.22cm}p{.22cm}c|p{.22cm}p{.22cm}c|p{.25cm}}

    \toprule
         \bf{Dataset 2} & \multicolumn{3}{c|}{\textbf{Ensemble (RF)}} & \multicolumn{3}{c|}{\textbf{Ensemble (LG)}} & \multicolumn{3}{c|}{\textbf{Ensemble (B)}} & \multicolumn{3}{c|}{\textbf{Ensemble (RB)}}\\
        Information Type & Pr & Rc & F1  & Pr & Rc & F1  & Pr & Rc & F1  & Pr & Rc & F1  & Sup \\ 
        \midrule
        Action on Issue & 0.67 & 0.79 & 0.72 & 0.72 & 0.78 & \textbf{0.74} & 0.69 & 0.77 & 0.71 & 0.55 & 0.82 & 0.65 & 96 \\ 
        Bug Reproduction & 0.63 & 0.54 & \textbf{0.58} & 0.65 & 0.49 & 0.55 & 0.53 & 0.59 & 0.53 & 0.5 & 0.61 & 0.54 & 176 \\ 
        Contribution \& Commit. & 0.71 & 0.6 & 0.62 & 0.76 & 0.6 & \textbf{0.64} & 0.61 & 0.62 & 0.59 & 0.54 & 0.64 & 0.55 & 109 \\ 
        Expected Behaviour & 0.38 & 0.17 & 0.22 & 0.38 & 0.17 & 0.23 & 0.16 & 0.25 & 0.18 & 0.23 & 0.37 & \textbf{0.26 }& 24 \\ 
        Investigation and Explr.  & 0.72 & 0.84 & \textbf{0.77} & 0.71 & 0.83 & 0.76 & 0.78 & 0.65 & 0.71 & 0.81 & 0.59 & 0.68 & 1891 \\ 
        Motivation & 0.42 & 0.33 & 0.35 & 0.54 & 0.44 & 0.47 & 0.83 & 0.56 & \textbf{0.65} & 0.61 & 0.5 & 0.52 & 18 \\ 
        Observed Bug Behav. & 0.39 & 0.3 & \textbf{0.33} & 0.49 & 0.23 & 0.3 & 0.18 & 0.25 & 0.2 & 0.21 & 0.37 & 0.26 & 81 \\ 
        Social Conversation & 0.72 & 0.53 & \textbf{0.61} & 0.67 & 0.56 & \textbf{0.61} & 0.61 & 0.62 & \textbf{0.61} & 0.59 & 0.62 & 0.6 & 469 \\ 
        Solution Discussion & 0.71 & 0.67 & \textbf{0.69} & 0.68 & 0.67 & 0.68 & 0.64 & 0.72 & 0.68 & 0.62 & 0.72 & 0.66 & 856 \\ 
        Task Progress & 0.53 & 0.37 & 0.43 & 0.5 & 0.36 & 0.41 & 0.4 & 0.45 & 0.42 & 0.45 & 0.49 & \textbf{0.46} & 233 \\ 
        Testing & 0.67 & 0.6 & \textbf{0.63} & 0.69 & 0.58 & 0.62 & 0.58 & 0.65 & 0.61 & 0.53 & 0.67 & 0.59 & 261 \\ 
        Workarounds & 0.67 & 0.63 & 0.61 & 0.76 & 0.59 & \textbf{0.63} & 0.57 & 0.63 & 0.56 & 0.62 & 0.63 & 0.59 & 27 \\ 
        \midrule
        \bf{Micro Average} & 0.69 & 0.69 & 0.69 & 0.69 & 0.69 & 0.69 & 0.64 & 0.64 & 0.64 & 0.62 & 0.62 & 0.62 & 4330 \\ 
        \bf{Macro Average} & 0.59 & 0.53 & 0.54 & 0.63 & 0.53 & 0.55 & 0.54 & 0.56 & 0.53 & 0.52 & 0.58 & 0.53 & 4330 \\ 
        \bf{Weighted Average} & 0.69 & 0.69 & \textbf{0.68} & 0.68 & 0.69 & 0.67 & 0.67 & 0.64 & 0.64 & 0.66 & 0.62 & 0.63 & 4241 \\ 
        \bottomrule
    \end{tabular}
    \end{adjustbox}
\end{table}

In Dataset 1, the ensemble classifiers generally achieve a weighted average F1-score ranging from 0.70 to 0.72, highlighting their consistent performance across diverse information types. Notably, the LightGBM ensemble stands out by achieving the highest F1-score of 0.72, and it notably excels in seven out of thirteen information categories, indicating its particular strength in extracting and classifying nuanced information effectively. Meanwhile,  in Dataset 2, the weighted average F1-scores for all models hover around 0.64 to 0.68, with Random Forest slightly leading with a 0.68 F1-score, reinforcing the adaptability of this model to different datasets.

\begin{table}[!t]
    \centering
    \caption{Performance of ensemble classifiers on each fold of dataset 1. (RF = Random Forest, LG = LightGBM, B = BERT, RB = RoBERTa)}
    \label{tab:fold_result_ensemble_d1}
    \begin{adjustbox}{width=\columnwidth}
    \centering
    \fontsize{6pt}{6pt}\selectfont
    \begin{tabular}{c|p{.4cm}p{.4cm}c|p{.4cm}p{.4cm}c|p{.4cm}p{.4cm}c|p{.4cm}p{.4cm}c}
    \toprule
 & \multicolumn{3}{c|}{\textbf{Ensemble RF}} & \multicolumn{3}{c|}{\textbf{Ensemble LG}} & \multicolumn{3}{c|}{\textbf{Ensemble B}} & \multicolumn{3}{c}{\textbf{Ensemble RB}} \\ 
        
Fold & Pr & Rc & F1 & Pr & Rc & F1 & Pr & Rc & F1 & Pr & Rc & F1 \\
\hline
1  & 0.70 & 0.70 & 0.70 & 0.67 & 0.67 & 0.67 & 0.69 & 0.69 & 0.68 & 0.73 & 0.72 & 0.72 \\
2  & 0.71 & 0.72 & 0.71 & 0.72 & 0.72 & 0.72 & 0.72 & 0.71 & 0.71 & 0.73 & 0.71 & 0.72 \\
3  & 0.72 & 0.71 & 0.71 & 0.71 & 0.69 & 0.69 & 0.71 & 0.71 & 0.70 & 0.72 & 0.68 & 0.69 \\
4  & 0.72 & 0.70 & 0.70 & 0.72 & 0.71 & 0.71 & 0.74 & 0.72 & 0.72 & 0.73 & 0.68 & 0.69 \\
5  & 0.73 & 0.73 & 0.72 & 0.74 & 0.73 & 0.72 & 0.73 & 0.72 & 0.72 & 0.72 & 0.70 & 0.70 \\
6  & 0.68 & 0.68 & 0.67 & 0.69 & 0.68 & 0.67 & 0.67 & 0.67 & 0.67 & 0.71 & 0.70 & 0.70 \\
7  & 0.75 & 0.75 & 0.75 & 0.73 & 0.71 & 0.72 & 0.73 & 0.73 & 0.73 & 0.76 & 0.74 & 0.75 \\
8  & 0.73 & 0.71 & 0.71 & 0.69 & 0.68 & 0.68 & 0.70 & 0.69 & 0.69 & 0.72 & 0.68 & 0.68 \\
9  & 0.77 & 0.76 & 0.75 & 0.75 & 0.75 & 0.74 & 0.74 & 0.73 & 0.73 & 0.76 & 0.75 & 0.75 \\
10 & 0.71 & 0.70 & 0.70 & 0.70 & 0.69 & 0.69 & 0.69 & 0.68 & 0.68 & 0.72 & 0.71 & 0.71 \\
\midrule
Mean     & 0.72 & 0.72 & 0.71 & 0.71 & 0.70 & 0.70 & 0.71 & 0.71 & 0.70 & 0.73 & 0.71 & 0.71 \\
Variance & 0.0006 & 0.0005 & 0.0005 & 0.0006 & 0.0006 & 0.0005 & 0.0005 & 0.0004 & 0.0004 & 0.0003 & 0.0005 & 0.0005 \\
\bottomrule
\end{tabular}
\end{adjustbox}
\end{table}
\begin{table}[!t]
    \centering
    \caption{Performance of ensemble classifiers on each fold of dataset 2. (RF = Random Forest, LG = LightGBM, B = BERT, RB = RoBERTa)}
    \label{tab:fold_result_ensemble_d2}
    \begin{adjustbox}{width=\columnwidth}
    \centering
    \fontsize{6pt}{6pt}\selectfont
    \begin{tabular}{c|p{.4cm}p{.4cm}c|p{.4cm}p{.4cm}c|p{.4cm}p{.4cm}c|p{.4cm}p{.4cm}c}
    \toprule
 & \multicolumn{3}{c|}{\textbf{Ensemble RF}} & \multicolumn{3}{c|}{\textbf{Ensemble LG}} & \multicolumn{3}{c|}{\textbf{Ensemble B}} & \multicolumn{3}{c}{\textbf{Ensemble RB}} \\ 
        
Fold & Pr & Rc & F1 & Pr & Rc & F1 & Pr & Rc & F1 & Pr & Rc & F1 \\
\hline
1  & 0.65 & 0.66 & 0.65 & 0.65 & 0.66 & 0.65 & 0.63 & 0.64 & 0.62 & 0.64 & 0.64 & 0.64 \\
2  & 0.64 & 0.67 & 0.65 & 0.66 & 0.68 & 0.66 & 0.65 & 0.67 & 0.66 & 0.63 & 0.64 & 0.63 \\
3  & 0.65 & 0.65 & 0.64 & 0.66 & 0.65 & 0.64 & 0.63 & 0.55 & 0.56 & 0.63 & 0.61 & 0.62 \\
4  & 0.72 & 0.72 & 0.71 & 0.73 & 0.74 & 0.73 & 0.71 & 0.67 & 0.68 & 0.70 & 0.66 & 0.67 \\
5  & 0.71 & 0.71 & 0.70 & 0.68 & 0.68 & 0.67 & 0.66 & 0.65 & 0.65 & 0.64 & 0.58 & 0.59 \\
6  & 0.72 & 0.73 & 0.72 & 0.67 & 0.66 & 0.66 & 0.73 & 0.69 & 0.70 & 0.69 & 0.59 & 0.60 \\
7  & 0.70 & 0.71 & 0.70 & 0.69 & 0.71 & 0.69 & 0.69 & 0.68 & 0.68 & 0.70 & 0.65 & 0.66 \\
8  & 0.73 & 0.71 & 0.71 & 0.69 & 0.69 & 0.68 & 0.65 & 0.63 & 0.63 & 0.71 & 0.65 & 0.66 \\
9  & 0.69 & 0.69 & 0.68 & 0.70 & 0.71 & 0.69 & 0.66 & 0.61 & 0.61 & 0.67 & 0.61 & 0.61 \\
10 & 0.67 & 0.67 & 0.66 & 0.68 & 0.68 & 0.67 & 0.65 & 0.61 & 0.62 & 0.64 & 0.59 & 0.60 \\
\midrule
Mean     & 0.688 & 0.692 & 0.682 & 0.681 & 0.686 & 0.674 & 0.670 & 0.640 & 0.640 & 0.670 & 0.620 & 0.630 \\
Variance & 0.0010 & 0.0007 & 0.0008 & 0.0005 & 0.0007 & 0.0006 & 0.0010 & 0.0016 & 0.0015 & 0.0009 & 0.0008 & 0.0007 \\
\bottomrule
\end{tabular}
\end{adjustbox}
\end{table}

With respect to recall, the ensemble models show notable improvements over stand-alone models for the most critical information types. On Dataset 1, Ensemble (RF) achieves strong recall for Observed Bug Behaviour (0.73), improving over the best stand-alone LightGBM recall of 0.69, and improves Workarounds recall from 0.44 (standalone LightGBM) to 0.60, meaning fewer workaround comments are missed. Ensemble (LG) achieves the highest recall for Bug Reproduction (0.69), improving over standalone LightGBM (0.62), and matches Solution Discussion recall (0.81) which is comparable to standalone LightGBM (0.82). On Dataset 2, Ensemble (RF) achieves the highest recall for Action on Issue (0.79) and Investigation and Exploration (0.84), while Ensemble (LG) leads on Investigation and Exploration recall (0.83). Solution Discussion recall improves from 0.60 (standalone LightGBM) to a range of 0.67--0.72 across ensemble models, reflecting consistent detection of solution-related discussions. Some information types such as Observed Bug Behaviour, Task Progress, and Expected Behaviour on Dataset 2 show lower recall, which is expected given their relatively small support in the dataset. Overall, ensemble models consistently improve recall for the most critical information types, reducing the risk of missing important information in issue discussions.

\subsubsection{Misclassification Analysis}\label{sec:misclassification}

\begin{table}[!t]
    \centering
    \caption{How the misclassification of the best stand-alone model can be corrected ensemble classifier}
    \label{tab:wrong_ensemble}
    \begin{tabular}{lp{1.5cm}p{1.5cm}p{1.5cm}p{1cm}}
    
        \toprule
        \bf{Information Type} & \bf{Ensemble (RF)} & \bf{Ensemble (LG)} & \bf{Ensemble (B)} & \bf{Ensemble (RB)} \\ 
        \midrule
        \multicolumn{5}{c}{\textbf{Dataset 1}}\\
        \midrule
        Action on Issue & 51.52\% & 63.64\% & 42.42\% & 54.55\% \\ 
        Bug Reproduction & 22.83\% & 20.65\% & 28.26\% & 41.30\% \\ 
        Contribution \& Commit. & 28.36\% & 28.36\% & 35.82\% & 56.72\% \\ 
        Expected Behaviour & 23.29\% & 19.18\% & 17.81\% & 32.88\% \\ 
        Investigation and Explr. & 20.77\% & 16.92\% & 25.38\% & 26.92\% \\ 
        Motivation & 25.32\% & 34.42\% & 31.82\% & 35.06\% \\ 
        Observed Bug Behav. & 14.63\% & 24.39\% & 14.63\% & 19.51\% \\ 
        Potential New Issues. & 14.86\% & 21.62\% & 21.62\% & 24.32\% \\ 
        Social Conversation & 43.48\% & 28.70\% & 45.65\% & 39.13\% \\ 
        Solution Discussion & 28.52\% & 25.78\% & 32.42\% & 15.63\% \\ 
        Task Progress & 25.71\% & 20.95\% & 35.24\% & 40.00\% \\ 
        Usage & 13.51\% & 24.32\% & 22.97\% & 33.78\% \\ 
        Workarounds & 16.00\% & 30.00\% & 26.00\% & 48.00\% \\ 

        \midrule
        \multicolumn{5}{c}{\textbf{Dataset 2}}\\
        \midrule
        Action on Issue & 62.96\% & 64.81\% & 64.81\% & 72.22\% \\ 
        Bug Reproduction & 19.05\% & 27.62\% & 35.24\% & 38.10\% \\ 
        Contribution \& Commit. & 41.43\% & 45.71\% & 45.71\% & 51.43\% \\ 
        Expected Behaviour & 13.64\% & 18.18\% & 27.27\% & 36.36\% \\ 
        Investigation and Explr. & 32.86\% & 37.09\% & 14.55\% & 7.98\% \\ 
        Motivation & 12.50\% & 0.00\% & 12.50\% & 0.00\% \\ 
        Observed Bug Behav. & 1.69\% & 3.39\% & 5.08\% & 16.95\% \\ 
        Social Conversation & 29.23\% & 25.00\% & 40.38\% & 40.38\% \\ 
        Solution Discussion & 33.33\% & 33.63\% & 43.66\% & 46.61\% \\ 
        Task Progress & 26.02\% & 29.59\% & 37.76\% & 42.35\% \\ 
        Testing & 25.98\% & 29.13\% & 36.22\% & 40.16\% \\ 
        Workarounds & 9.09\% & 18.18\% & 18.18\% & 18.18\% \\ 
        \bottomrule

    \end{tabular}
\end{table}

Table \ref{tab:wrong_ensemble} presents correction rates across various categories for both Dataset 1 and Dataset 2 using ensemble classifiers. Within each category, Each value represents the percentage corrections over the best stand-alone model. Each ensemble classifier shows varying levels of corrections in different categories, reflecting their strengths and weaknesses. For example, in Dataset 1, Contribution and Commitment sees substantial corrections with RoBERTa, achieving a 56.72\% improvement, suggesting its effectiveness in understanding nuanced textual data in this category. RoBERTa and BERT often lead in terms of corrections percentages of the stand-alone LightGBM model.
\begin{tcolorbox}[flushleft upper, boxrule=1pt, arc=0pt, left=0pt, right=0pt, top=0pt, bottom=0pt, colback=white, after=\ignorespacesafterend\par\noindent]
\textbf{Summary of RQ2.} The ensemble model achieves the highest F1-score for information type detection in both datasets, surpassing SOTA model~\cite{InfoTypesOSSDiscussion} by 11-13\%. 
\end{tcolorbox}

\section{Effectiveness of iSum}\label{sec:effectiveness}
In this section, we aim to assess the effectiveness of iSum through an empirical user study focused on the task of issue understanding. Participants will engage in tasks involving the use of iSum and compare their experiences with and without the tool. Based on the results, we answer the following research questions:

\begin{enumerate}[label=\bf{RQ\arabic{*}.}, start=3]
    \item How effective is iSum to support developers during bug report understanding?
    \item Can iSum also help to explore feature enhancement requests in issue reports?
\end{enumerate}

\subsection{Participants and Data Collection}\label{sec:datacollection}
In this user study, we recruited 30 participants, predominantly comprised of developers (27 participants), alongside QA engineers (2 participants) and a project manager (1 participant). These participants were recruited through personal connections. Their experience in the software industry varied, with 50\% having between 4 and 5 years of experience and 14 participants with 0 to 3 years. Each participant was assigned to one of two groups, with each group receiving a different arrangement of tasks involving the use of iSum. The study did not have participants perform two iterations on the same issue. Group 1 analyzed a bug report without the aid of iSum and a feature proposal with iSum, while Group 2 received the inverse setup—employing iSum for the bug report and proceeding without it for the feature proposal. This structure was chosen to distinguish the effects of iSum on both the speed and quality of information extraction and comprehension across various types of issue discussions. Both groups contain an equal number (15) of participants. The study was conducted using Google Forms, with separate forms for each group. Participants initially engaged in manual understanding of an issue report and rated the perceived difficulties faced during this process, followed by another issue understanding with iSum and rating the usefulness of the tool. Prior to using iSum, they received a guide explaining each information type and how the tool functions. To prevent fatigue, participants were encouraged to take short breaks between tasks. In addition to completing coding tasks, participants were asked to assess the complexity and load required for each task using the NASA Task Load Index (TLX) \cite{nasatlx}, which evaluates subjective workload. Participants gave free-form textual answers to task questions like "What bug behavior was observed?" or "What solutions were discussed?" They weren't asked to classify individual sentences—just respond naturally. Authors manually identified relevant sentences for each info type.

\subsection{Variables to Assess Task Performance} \label{sec:nasatlx}
To measure the effectiveness of iSum in facilitating issue report understanding, we designed a set of dependent variables that capture various aspects of task performance. These metrics allow us to quantitatively assess the improvements provided by iSum in comparison to traditional methods of issue report understanding.

\bf{Correctness.} This metric evaluates how accurately participants identified the classified information types within the issue discussions. Correctness (C) was assessed by comparing participants' responses against a predefined set of answers. To establish the ground truth, the same author who labeled Dataset 2 conducted a thorough manual analysis of the issue discussions. Sentences related to the needed information type are labeled and periodically reviewed for consistency and accuracy against the established definitions of information types. Upon this, candidate answers have been created. Before similarity analysis, each response is manually segmented under each information type, and for information types where multiple answers may exist, the response is deemed correct if it successfully identifies one or several of the expected answers. The Correctness score (C) is calculated using the following formula:

\begin{equation}\label{eq:correctness}
    C = \frac{\text{Number of Correctly Identified Information Types}}{\text{Expected Number of Information Types}} 
\end{equation}

To automate the evaluation of responses, we utilize the SentenceTransformer library, employing the allMiniLM-L6-v2 model \cite{huggingface} to encode the actual and expected answers into embeddings. Subsequently, we gauge the semantic similarity between these embeddings by computing the cosine similarity to compare against the threshold of 0.7 \cite{Rahutomo2012SemanticCS} and assessing how closely the actual answer aligns with the expected summary. For example, in the TensorFlow \#33030 memory leak task asking “What solutions were discussed?”, our manually curated ground-truth exemplar was: “Calling \texttt{tf.keras.backend.clear\_session()} after every \texttt{model.predict()} fixes the issue” (from issue comments). A participant responded: “After using \texttt{tf.keras.backend.clear\_session()}, the memory leak is fixed.” The SentenceTransformer cosine similarity exceeded the $0.7$ threshold, marking the response as correct for the \textit{Solution Discussion} information type and awarding full task credit—demonstrating semantic matching rather than verbatim reproduction.


\bf{Task Load.} We used the NASA TLX metrics values for computing the task loads of participants. Participants rated the task load required on a scale from 1 to 100, where higher values indicate more task load. This self-reported metric reflects the cognitive and physical demand of the task, with and without iSum. Six aspects were assessed :
\begin{inparaenum}
\item {Mental Demand:} The extent of mental and perceptual activity required
\item {Physical Demand:} The level of physical activity involved
\item {Temporal Demand:} The pressure of time constraints during task execution
\item {Performance:} The success in achieving task goals and satisfaction with the outcomes
\item {Effort:} The overall effort needed to achieve the performance level
\item {Frustration Level:} The emotional state experienced, ranging from stress to contentment.
\end{inparaenum}
Finally, the overall TLX score is the average of each participant's TLX score, $TLX_i$:

\begin{equation}
    TLX_i = \frac{\sum_{j=1}^{6} S_{i,j}}{6}
\end{equation}

$\quad S_{i,j} = \text{score for dimension } j \text{ given by participant } i$

\subsection{Statistical Analysis} \label{sec:statisticalAnalysis} To assess the differences between the different settings in terms of correctness and task load, we employed the Mann-Whitney U test and Cliff’s Delta. 

\bf{Mann-Whitney U test.} The Mann-Whitney U test \cite{utest} was chosen for its suitability in analyzing non-normally distributed data, a common characteristic in experimental data in human factors research. This test evaluates whether two independent samples were selected from populations having the same distribution. Specifically, it was used to determine if the differences in Correctness scores and Task Load ratings between the two settings were statistically significant. A significance level (p-value) of less than 0.05 was considered indicative of a statistically significant difference between the two groups.
 
\bf{Cliff’s Delta.} This non-parametric measure of effect size determines the degree to which two distributions differ.  This metric ranges from -1 to 1, where values close to 0 suggest similar distributions, and values approaching -1 or 1 indicate greater disparity. To interpret Cliff’s Delta, we adopt thresholds based on the recommendations by Romano et al \cite{romano2006exploring}. 

\text{Effect Size} = 
$\begin{cases}
    \text{Negligible (N)}, & \text{if } |\delta| \leq 0.147 \\
    \text{Small (S)}, & \text{if } 0.147 < |\delta| \leq 0.33 \\
    \text{Medium (M)}, & \text{if } 0.33 < |\delta| \leq 0.474 \\
    \text{Large (L)}, & \text{if } 0.474 < |\delta| \leq 1 \\
\end{cases}$

\subsection{Effectiveness of iSum in Bug Report Understanding (RQ3)} \label{sec:rq3}
\subsubsection{Study Design} In the task design section of this user study, we selected a bug report, issue \#33030  from TensorFlow GitHub repository to evaluate the effectiveness of our tool, iSum, in navigating complex discussion threads. To ensure a focused and time-efficient assessment, each participant was instructed to complete their assigned tasks within a strict five-minute window. This setup aimed to simulate real-world conditions where quick issue resolution is often necessary.  \tbl\ref{tab:task_details} presents the details of the task setting. Participants were assigned the role of a software developer and provided with a bug report that included a lengthy and complex discussion thread. Their task was to identify the bug's behavior as described in the report, investigate potential causes mentioned by stakeholders, if any, and identify any proposed solutions or workarounds discussed. This task aimed to evaluate participants' ability to extract these four types of information from issue reports within a 5-minute time frame.  \tbl\ref{tab:task_details} presents the related information types for this task. Half of the participants performed the task with the assistance of iSum (BI), while the other half did it without (BM). The performance difference in finding the correct information were used to assess iSum's effectiveness in facilitating quick and thorough understanding of technical issues from complex information threads, which is critical for timely and effective bug resolution.

\begin{table}[!t]
\centering
\caption{Task details. Here, G\# = Group \#, Feat. = Feature request}
\label{tab:task_details}
    \begin{adjustbox}{width=\columnwidth}
    \centering
    \fontsize{6pt}{6pt}\selectfont
\begin{tabular}{p{5mm} p{5mm} p{5mm} p{3mm} p{40mm} p{35mm}}
\toprule
\textbf{\newline Task} & \textbf{Issue Type} & \textbf{iSum Used} & \textbf{\newline G\#} & \textbf{\newline Task Details} & \textbf{\newline Related Information Type} \\ 
\midrule
BM & Bug & No & G1 & \multirow{3}{40mm}{Investigate the bug's behavior and possible causes. 
Look for proposed solutions or workarounds.} 
& \multirow{3}{35mm}{Obs. Bug Behaviour, Investigation. , 
Solution Discussion, Workaround} \\ 
BI & Bug & Yes & G2 & & \\ 
\\
\midrule
FM & Feat. & No & G2 
& \multirow{6}{40mm}{Understand the motivations behind the proposal and the expected outcomes.
Look for contribution interest or assignments, new challenges or feature requests and 
monitor the task progress} 

& \multirow{6}{35mm}{Motivation, Expected Behaviour, Potential New Issues and Requests , Contribution and Commitment, Task Progress} \\ 
FI & Feat. & Yes & G1 & & \\ 
\\
\\
\\

\bottomrule
\end{tabular}
\end{adjustbox}
\end{table}

\subsubsection{Study Results}\label{sec:rq3Results}
\customTwoBarChart{73.3} {-20} {-6.7}  {66.7} {-13.3} {-20}{bug report understanding}{fig:T1}

The feedback from participants as illustrated in \fig\ref{fig:T1}, indicates that extracting relevant information and understanding the overall context of an issue is challenging without the aid of information type classification. Most participants found it difficult, with 11 rating it as high difficulty. On the other hand, the effectiveness of iSum in aiding this process received mostly positive feedback. Out of the participants, 10 reported a positive impact, suggesting that iSum facilitated the extraction and understanding of relevant information within issue discussions. Meanwhile, 2 participants felt neutral about its effectiveness, and 3 experienced a negative impact, indicating variability in how iSum's classifications met individual needs or expectations in understanding issue reports.

\usepgfplotslibrary{statistics}
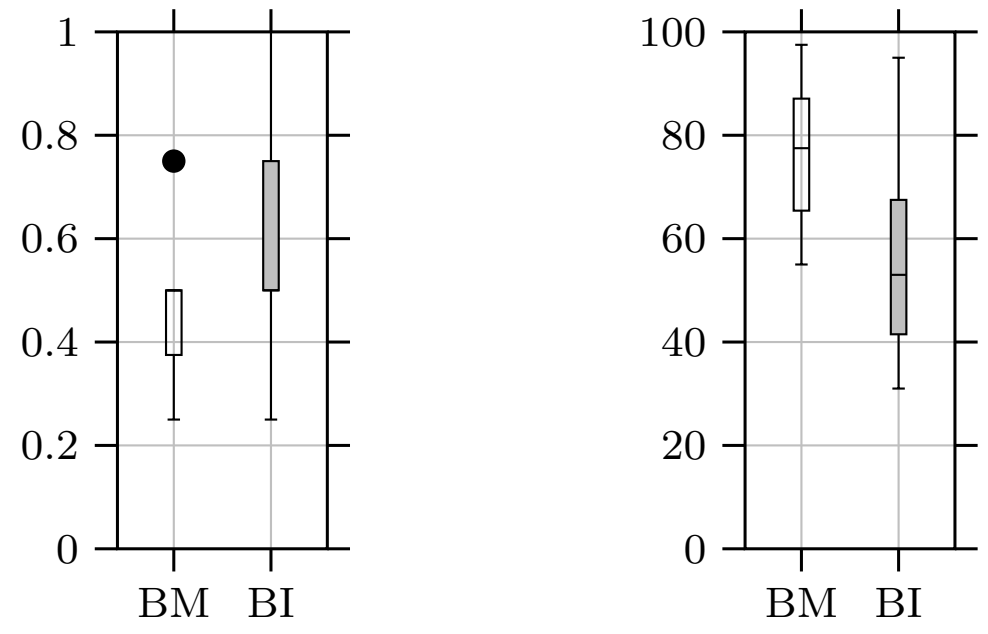
\begin{figure}[!t]
    \centering
    \begin{subfigure}
    \centering
    \begin{tikzpicture}
    \begin{axis}[
                boxplot/draw direction=y,
                boxplot/variable width,
                ymin=0,
                ymax=1,
                ytick={0,0.2,0.4,0.6,0.8,1},
                axis line style={-},
                xticklabels={BM, BI},
                xtick={1,2},
                axis line style={black,line width=0.2mm},
                xtick align=outside,
                ytick align=outside,
                ymajorgrids=true,
                xmajorgrids=true,
                tick style={black,line width=0.2mm}, 
                width=0.25\textwidth, 
                height=5cm, 
                ]
    
      \addplot[boxplot] table[row sep=\\,y index=0] {
        0.5 \\ 0.5 \\ 0.5 \\ 0.5 \\ 0.5 \\ 0.25 \\ 0.25 \\ 0.5 \\ 0.5 \\ 0.5 \\ 0.75 \\ 0.5 \\ 0.5 \\ 0.25 \\ 0.25 \\
      };
    
        \addplot[boxplot, fill=gray!50] table[row sep=\\,y index=0] { 
        0.75 \\ 0.75 \\ 1 \\ 0.5 \\ 1 \\ 0.75 \\ 0.5 \\ 0.5 \\ 0.75 \\ 0.5 \\ 0.5 \\ 0.5 \\ 0.75 \\ 0.5 \\ 0.25 \\
      };
    \end{axis}
    \end{tikzpicture}
    \end{subfigure}
\hspace{1.5cm}
    \begin{subfigure}
    \centering
            \begin{tikzpicture}
    \begin{axis}[
                boxplot/draw direction=y,
                boxplot/variable width,
                ymin=0,
                ymax=100,
                ytick={0,20,40,60,80,100},
                axis line style={-},
                xticklabels={BM, BI},
                xtick={1,2},
                axis line style={black,line width=0.2mm},
                xtick align=outside,
                ytick align=outside,
                ymajorgrids=true,
                xmajorgrids=true,
                tick style={black,line width=0.2mm}, 
                width=0.25\textwidth, 
                height=5cm, 
                ]
    
      \addplot[boxplot] table[row sep=\\,y index=0] {
        55.00 \\ 63.33 \\ 67.50 \\ 70.00 \\ 85.00 \\ 97.50 \\ 92.50 \\ 60.00 \\ 95.00 \\ 87.50 \\ 86.67 \\ 77.50 \\ 79.17 \\ 77.50 \\ 56.67\\
      };
    
        \addplot[boxplot, fill=gray!50] table[row sep=\\,y index=0] { 
        40.00 \\ 43.00 \\ 58.00 \\ 95.00 \\ 39.00 \\ 65.00 \\ 73.00 \\ 90.00 \\ 49.00 \\ 68.00 \\ 35.00 \\ 67.00 \\ 31.00 \\ 53.00 \\ 49.00\\
      };
    \end{axis}
    \end{tikzpicture}
    \end{subfigure}
    \caption{(a) Correctness of the task and (b) Task load spent across settings for bug report understanding. Here, BI = Bug report understanding with iSum and BM = without iSum.}
    \label{fig:BT_boxplots}

\end{figure}

In \fig\ref{fig:BT_boxplots}, we show the correctness of the tasks related to bug report understanding and task load across the two settings (manual analysis and analysis using iSum) using two box plots, one for each setting. In comparing the correctness of information type identification between manual analysis and the use of iSum, the data reveals that the mean correctness score increased from 0.45 (BM) to 0.63 (BI) for bug report understanding after using iSum and enhances participants' ability to correctly identify information. For instance, while the solution to the problem of training for one epoch at a time, saving the model, and reloading it in subsequent iterations was buried within a heap of comments, iSum was able to effectively surface this solution.  One participant noted, \textit{"easier to find what I need,"} highlighting the tool's effectiveness in locating pertinent details.  However, the median for both BM and BI is 0.5. The improvement with iSum is therefore more visible in the mean than the median, reflecting that iSum helped more participants achieve higher correctness scores. Furthermore, when participants explored bug reports manually, the median task load was significantly higher at 77.50 compared to 53.00 when using iSum. This highlights iSum's effectiveness in reducing the cognitive load and time required to navigate complex bug reports. 

\begin{table}[!t]
\centering
\caption{Results of Wilcoxon Mann-Whitney U test and Cliff's $\delta$ effect size for the pair-wise comparisons of the correctness and task load on bug understanding task without and with the assistance of iSum (BM vs BI) }
\label{tab:effect_bug}
\begin{tabular}{lll}
\toprule
\textbf{Metric} & \textbf{\textit{p-value}} & \textbf{\textit{Cliff's $\delta$}} \\ \midrule
Correctness & 0.012 & -0.489 L\\
Average TLX & 0.007 & 0.587 L\\ 
\midrule
Mental Demand & 0.003 & 0.640 L \\
Physical Demand & 0.573 & 0.124 N \\
Temporal Demand & 0.001 & 0.698 L \\
Performance & 0.028 & 0.471 M \\
Effort & 0.001 & 0.711 L \\
Frustration Level & 0.004 & 0.622 L \\ \bottomrule
\end{tabular}

\end{table}

In \tbl\ref{tab:effect_bug}, we present the statistical results from the Wilcoxon Mann-Whitney U tests and Cliff's $\delta$ calculations to show the comparison between the Manual method and iSum across different metrics of bug report understandings. The p-value of 0.012 indicates a statistically significant difference in correctness between Manual and iSum, with a Cliff's $\delta$ of -0.489. This suggests that iSum significantly improves correctness in bug report understanding compared to the manual method, with a large effect size indicating substantial improvement. Again, the p-value of 0.007 suggests a significant difference in overall task load, with a very high Cliff's $\delta$ of 0.587. This indicates that iSum greatly reduces the task load required in bug report understanding, showing a strong effect where iSum is much more efficient than the manual approach. Furthermore, the performance dimension shows a Cliff's $\delta$ of 0.471, which is classified as Medium (M) according to the thresholds of Romano et al. \cite{romano2006exploring}, where a Large effect requires $|\delta| > 0.474$. However, the value is borderline M/L, only 0.003 below the Large threshold. The relatively higher p-value (0.028 vs., for example, 0.001 for Temporal Demand) likely reflects greater variability in how participants perceived their own performance satisfaction. Even when iSum helped participants locate information more efficiently, some participants remained uncertain about the completeness of their answers, resulting in more variable self-reported performance scores compared to more objective dimensions such as Effort or Temporal Demand. Nevertheless, the result remains statistically significant ($p < 0.05$) with a near-large effect size, indicating that iSum meaningfully reduced perceived performance load.\\

\begin{tcolorbox}[flushleft upper, boxrule=1pt, arc=0pt, left=0pt, right=0pt, top=0pt, bottom=0pt, colback=white, after=\ignorespacesafterend\par\noindent]
\textbf{Summary of RQ3.} 
Participants found it challenging to extract relevant information and understand the overall context of an issue without the aid of information type classification and iSum positively impacted the process for most participants. The use of iSum resulted in increased correctness of information type identification and reduced cognitive load required to navigate complex bug report compared to manual analysis.
\end{tcolorbox}

\subsection{Effectiveness of iSum in Feature Request Understanding (RQ4)} \label{sec:rq4}
\subsubsection{Study Design}
For this task, we selected a feature request, issue \#206, from TensorFlow and also gave participants a five-minute window to complete understanding task as detailed in \tbl\ref{tab:task_details}. In this task, participants assumed the role of a project manager reviewing a feature proposal. Again, this challenge involved delving into a detailed discussion thread related to the feature, understanding the motivations behind the proposal and the outcomes it aimed to achieve. Furthermore, they need to look for any potential challenges or additional feature requests that may discuss in the comment section; also monitor any sign of enthusiasm, willingness to contribute, or preliminary task assignments within the discussion. Lastly, they need to monitor the progress of tasks. For this feature understanding also, there are two task settings, understanding with (FI) or without iSum (FM) and extracting key insights relevant to the feature proposal. Both studies did not have participants perform two iterations on the same issue. We employed a between-subjects design: G1 analyzed the bug report manually (BM) and the feature report with iSum (FI), while G2 analyzed the bug report with iSum (BI) and the feature report manually (FM). Therefore, each participant encountered each issue only once.

\subsubsection{Study Results}\label{sec:rq4Results}
\customTwoBarChartIsumAnother{93.3} {-6.67} {0}  {40} {-40} {-20}{feature report understanding}{fig:T2}
In the user study focused on feature report understanding, the response for evaluating the difficulty of extracting relevant information and understanding the overall context without the aid of information type classification was overwhelming. This is illustrated in \fig\ref{fig:T2}, with 14 out of 15 participants rating the difficulty as high. For assessing the effectiveness of the iSum classifications for feature report understanding, participants provided mixed feedback. Out of 15 participants, 6 reported a positive impact, stating that iSum classifications aided in extracting relevant information and understanding the overall context of the issue. Another 6 participants remained neutral, indicating neither significant benefit nor drawback from using iSum. Remaining three participants rated iSum as “Not useful” mainly because of misclassified or missing tags and perceived mismatches between the tool’s labels and what they needed for nuanced feature discussions (e.g., side questions not tagged as Potential New Issues and Requests).

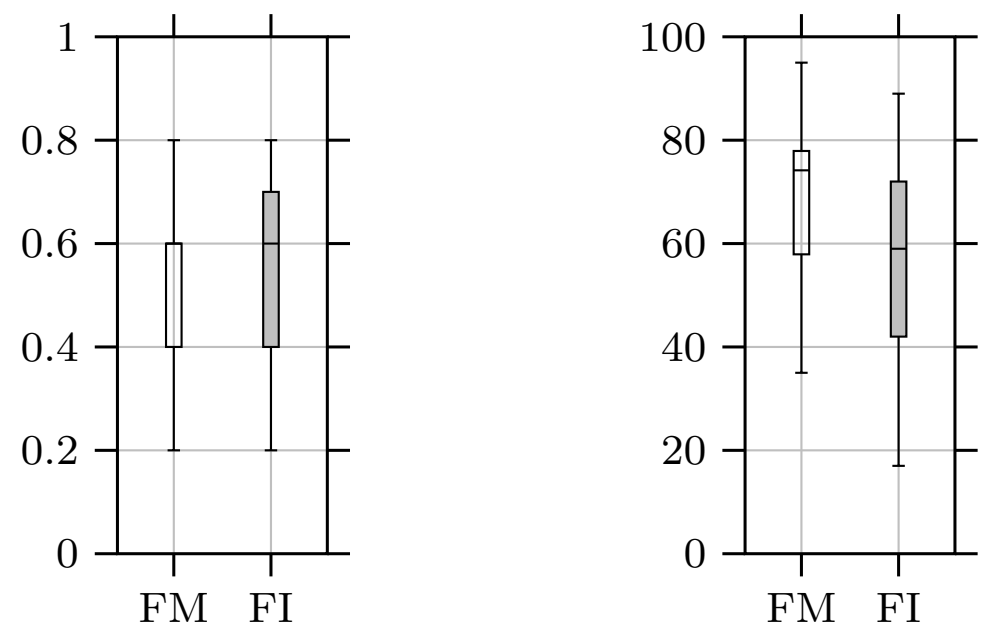
\begin{figure}[h]
    \centering
    \begin{subfigure}
    \centering
    \begin{tikzpicture}
    \begin{axis}[
                boxplot/draw direction=y,
                boxplot/variable width,
                ymin=0,
                ymax=1,
                ytick={0,0.2,0.4,0.6,0.8,1},
                axis line style={-},
                xticklabels={FM, FI},
                xtick={1,2},
                axis line style={black,line width=0.2mm},
                xtick align=outside,
                ytick align=outside,
                ymajorgrids=true,
                xmajorgrids=true,
                tick style={black,line width=0.2mm}, 
                width=0.25\textwidth, 
                height=5cm, 
                ]
    
      \addplot[boxplot] table[row sep=\\,y index=0] {
        0.6 \\ 0.6 \\ 0.4 \\ 0.6 \\ 0.4 \\ 0.2 \\ 0.4 \\ 0.2 \\ 0.4 \\ 0.6 \\ 0.6 \\ 0.6 \\ 0.6 \\ 0.8 \\ 0.4 \\
      };
    
        \addplot[boxplot, fill=gray!50] table[row sep=\\,y index=0] { 
        0.4 \\ 0.8 \\ 0.4 \\ 0.8 \\ 0.6 \\ 0.2 \\ 0.6 \\ 0.8 \\ 0.6 \\ 0.4 \\ 0.4 \\ 0.8 \\ 0.2 \\ 0.6 \\ 0.6 \\
      };
    \end{axis}
    \end{tikzpicture}
    \end{subfigure}
\hspace{1.5cm}
    \begin{subfigure}
    \centering
            \begin{tikzpicture}
    \begin{axis}[
                boxplot/draw direction=y,
                boxplot/variable width,
                ymin=0,
                ymax=100,
                ytick={0,20,40,60,80,100},
                axis line style={-},
                xticklabels={FM, FI},
                xtick={1,2},
                axis line style={black,line width=0.2mm},
                xtick align=outside,
                ytick align=outside,
                ymajorgrids=true,
                xmajorgrids=true,
                tick style={black,line width=0.2mm}, 
                width=0.25\textwidth, 
                height=5cm, 
                ]
    
            \addplot[boxplot] table[row sep=\\,y index=0] {
                75.00 \\ 70.00 \\ 80.83 \\ 75.00 \\ 90.83 \\ 75.00 \\ 56.67 \\ 74.17 \\ 42.50 \\ 95.00 \\ 52.50 \\ 59.17 \\ 35.00 \\ 91.67 \\ 62.50 \\
            };
            \addplot[boxplot, fill=gray!50] table[row sep=\\,y index=0] {
                31.00 \\ 17.00 \\ 34.00 \\ 68.00 \\ 50.00 \\ 78.00 \\ 89.00 \\ 25.00 \\ 75.00 \\ 69.00 \\ 68.00 \\ 58.00 \\ 59.00 \\ 84.00 \\ 52.00 \\
            };
    \end{axis}
    \end{tikzpicture}
    \end{subfigure}
    \caption{(a) Correctness of the task and (b) Task load spent across settings for feature report understanding. Here, FI = Feature request understanding with iSum and FM = without iSum.}
    \label{fig:ft_boxplots}

\end{figure}

The correctness and task load  spent for the tasks related to feature report understanding are shown in \fig\ref{fig:ft_boxplots}. In this task, iSum only slightly improves the mean correctness (0.49 (FM) to 0.55 (FI)) and its impact on the challenging scenarios was limited. For instance, during feature report understanding, one participant raised a side question regarding the optimal approach for achieving a specific type of indexing. But iSum failed to tag it as Potential Issues and Requests related comment. Consequently, participants relying on iSum may have missed this information, as highlighted by one user's comment: \textit{"Wrong tags make users lose information."} This indicates that iSum's performance may be hindered when discussions deviate from conventional issue-related topics or lack explicit tags. However, for the understanding of feature reports in the user study, the task load  required also saw a decrease when using iSum compared to manual methods; underscoring  the reduction in workload and task load  that iSum offers when managing and understanding feature reports. The median task load  score for manual understanding was 74.17, which dropped to 59.00 with the use of iSum. 

Again, the statistical results from the Wilcoxon Mann-Whitney U tests and Cliff's $\delta$ calculations comparing the Manual method and iSum across various metrics of feature report understandings are presented in Table \ref{tab:effect_feat}. With a p-value of 0.445, there is no statistically significant difference in correctness between the two methods for feature report understanding. However, the Cliff's $\delta$ of -0.160, classified as small (S), points towards a slight tendency for the manual method (FM) to potentially perform better in terms of correctness. Additionally, the p-value of 0.119 indicates no significant difference in task load, yet the high Cliff's $\delta$ of 0.320, indicates a small effect size favoring a reduction in task load when using iSum. This suggests that iSum potentially reduces task load  considerably in feature report understanding, although the results are not statistically significant.

\begin{table}[!t]
\centering
\caption{Results of Wilcoxon Mann-Whitney U test and Cliff's $\delta$ effect size for the pair-wise comparisons of the correctness and task load  on feature report understanding task with and without the assistance of iSum (FM vs FI)}
\label{tab:effect_feat}
\begin{tabular}{lll}
\toprule
\textbf{Metric} & \textbf{\textit{p-value}} & \textbf{\textit{Cliff's $\delta$}} \\ \midrule
Correctness & 0.445 & -0.160 S \\
Average TLX & 0.119 & 0.320 S \\ 
\midrule

Mental Demand & 0.139 & 0.320 S \\
Physical Demand & 0.298 & 0.227 S \\
Temporal Demand & 0.195 & 0.280 S \\
Performance & 0.133 & 0.324 S \\
Effort & 0.156 & 0.307 S \\
Frustration Level & 0.219 & 0.267 S \\ \bottomrule

\end{tabular}

\end{table}

\begin{tcolorbox}[flushleft upper, boxrule=1pt, arc=0pt, left=0pt, right=0pt, top=0pt, bottom=0pt, colback=white, after=\ignorespacesafterend\par\noindent]
\textbf{Summary of RQ4.} 
Feedback on the effectiveness of iSum for feature request understanding was mixed, with some participants reporting positive impacts on information extraction and understanding, while others remained neutral. Although iSum slightly improved the mean correctness of tasks related to feature report understanding, its impact on challenging scenarios was limited. Nevertheless, using iSum resulted in reduced task load  and workload compared to manual understanding methods.
\end{tcolorbox}

\section{Usability and Feasibility of iSum} \label{sec:feasibility}
In this section, we extend our user study with 10 industrial and OSS software engineers to investigate the practical usability, feasibility, and potential enhancements of iSum in real-world software development contexts. We show them this tool, explain its features and usage. By soliciting feedback and insights from industrial developers, the study sought to assess how well iSum aligns with the needs, preferences, and workflows of professionals working in diverse software development environments. Additionally, the study aimed to identify potential areas for improvement and refinement to ensure that iSum meets the evolving needs of its users and remains a valuable asset in the software development toolkit.  We aim to address the following research question:

\begin{enumerate}[label=\bf{RQ\arabic{*}.}, start=5]
    \item How do industrial developers perceive the usefulness of iSum?
\end{enumerate}
We divide this research question into two sub questions as follows:
\begin{enumerate}[label=\bf{RQ5.\arabic{*}.}, start=1, leftmargin=30pt]
    \item How do the developers consider of the usability of iSum?
    \item How do the developers consider of the feasibility of iSum in their daily tasks? 
\end{enumerate}

\subsection{Study Methodology}\label{sec:feasibilityStudyMethodology}
In this study, we conducted another series of semi-structured interviews with the same 10 participants from the previous phase (refer to Section \ref{sec:interview1}). This group comprised 10 professional software engineers, including six developers, two QA engineers, and two project managers. They bring varied levels of experience to the table, actively working within the software industry for several years. These interviews were conducted and recorded via Microsoft Teams, using automated transcription for data capture. The discussions focused on two primary topics: the usability and feasibility of iSum. The interview questions were a mix of closed and open-ended formats. Q1 and Q3 used binary yes/no response options, while Q2 was administered as a multi-select question with a predefined list of options. Q4-Q6 are open-ended questions. For closed-ended questions, responses were quantified, while an open-coding approach was applied to open-ended questions. We conducted open coding of each interview using an online qualitative data analysis software, Ligre \cite{LiGRESoftware}. This involved systematically reviewing the interview transcripts, identifying recurring themes, and assigning relevant codes to segments of text to categorize and organize the data.  Table \ref{tab:Feasibilityquestions} outlines the interview questions.

 \begin{table}[!t]
\caption{Interview Questions: Understanding Usability and Feasibility of iSum (C/O=Close/Open question)}
 \vspace{-3mm}
    \centering
    \begin{tabular}{c l c c}
     \toprule
     \bf{Q\#} & \bf{Questions} & \bf{O/C}\\ 
     \midrule
     \multicolumn{3}{c}{\cellcolor{gray!25}Usability (RQ6.1) }\\
     1 & Do you find the iSum tool useful? & C\\
     2 & Who do you think can benefit from the tool? & C\\
     3 & Will you or your team use this tool? & C\\
     \midrule
     \multicolumn{3}{c}{\cellcolor{gray!25}Feasibility  (RQ6.2)}\\
     4 & What are the strengths of this tool? & O\\
     5 & What are the shortcomings of this tool? & O\\
     6 & Do you have any suggestions for improving this tool? & O\\
     
     \bottomrule
    \end{tabular}
    \label{tab:Feasibilityquestions}
\end{table}

\subsection{Usability of iSum (RQ5.1)}

\subsubsection{Perceived Usefulness of iSum}\leavevmode

\noindent{80\%\it{ Participants perceived iSum as useful for both project and issue level summarization}~\color{black!90}\rule{48pt}{6pt}} \\
{20\%\it{ Participants perceived iSum as useful for only project level summarization}~\color{black!50}\rule{12pt}{6pt}} \\
Participants unanimously found the iSum tool to be useful, yet their appreciation of its features varied depending on their roles within the project. Two participants highlighted that summarization at the project level was particularly valuable since they found it best suited for understanding and managing broader project dynamics. Participant P02 articulated this viewpoint by stating, \it{"I mean from a developer perspective, because I believe in details and I believe developers should look into the details. But from a management perspective, this is awesome."} - P02. In contrast, the majority felt that both project and issue-level summarizations were beneficial, suggesting a wider applicability across different stakeholder needs. 


\subsubsection{Probable Beneficiaries  of iSum} \leavevmode

\noindent{80\%\it{ Participants consider project manager as a beneficiary of iSum}~\color{black!90}\rule{48pt}{6pt}} \\
{70\%\it{ Participants consider developer as a beneficiary of iSum} ~\color{black!50}\rule{42pt}{6pt}}\\
{20\%\it{ Participants consider newbie as a beneficiary of iSum}~\color{black!20}\rule{12pt}{6pt}}\\
Participants think this tool would be beneficial for three kinds of stakeholders: developers, project managers and anyone who is new to the project. Since developers are often confronted with the challenge of reading through extensive discussions to uncover essential details that impact their work, participants think this process can be frustrating and time-consuming and lead to delays and inefficiencies in issue resolution. P04 encapsulated this struggle, remarking, "Sometimes while developing, going through all the discussion is a bit frustrating or takes a lot of time. So it will benefit the developers." - P04. Conversely, P09 emphasized the daily importance of tracking the project's development and how iSum can assist with that: \it{"It's more for the project manager to understand the project on the daily basis."} - P09. Participants also noted that new team members, who may not be familiar with the project's history or specific issue threads, can use iSum to quickly understand the project's current state.  P09 explained the advantage clearly, \it{"Those who new to the project, doesn't know anything about the project; they can quickly get some information, some insight from the tool to understand what's going on."} - P09.

\subsubsection{Willingness to use iSum}\leavevmode

\noindent{90\%\it{ Participants are willing to use iSum}~\color{black!90}\rule{54pt}{6pt}}\\
{10\%\it{ Participants are not willing to use iSum} ~\color{black!50}\rule{6pt}{6pt}} \\
A significant majority (90\%) of participants expressed willingness to adopt the iSum tool, highlighting its potential value across multiple roles.  One added, they might like to use it as a base and personalize according to their need. Only one participant (P02) did not want to use it since, at that time, it lacked personalization options. We added personalization options after this interview and conducted the remaining eight interviews.

\subsection{Feasibility of iSum (RQ5.2)}

\subsubsection{Strengths of iSum}\leavevmode

\noindent{80\% \it{ Participants think iSum saves time in finding information}~\color{black!90}\rule{48pt}{6pt}} \\
{40\% \it{ Participants think iSum can provide an initial overview of the project/issue} ~\color{black!50}\rule{24pt}{6pt}} \\
Eight participants think the tool's success is in saving time in finding information from issue discussion. They preferred the highlighting and summary features the most. The customisation was also a favourite because everyone has various preferences depending on their role and practice. Four of them remarked on how helpful the tool is for getting a quick overview of the project or issue. One participant encapsulates both advantages in the following statement: \it{"The person concerned are getting information within few clicks and that is beneficial for developers. They can go to the specific issue and check the comment box or discussion and for project managers, they can use their filters, then create brief overview of his project."}-P07.

\subsubsection{Shortcomings of iSum}\label{sec:shortcoming}\leavevmode

\noindent{60\% \it{Participants consider inaccuracy of information detection as a shortcoming of iSum}~\color{black!90}\rule{36pt}{6pt}} \\
{40\% \it{ Participants think iSum needs integration of more information}~\color{black!50}\rule{24pt}{6pt}} \\
{10\% \it{ Participants think iSum may not help assigned developers}~\color{black!50}\rule{6pt}{6pt}} \\
{10\% \it{Participants faces difficulty in understanding information type}~\color{black!50}\rule{6pt}{6pt}} \\
Six participants are concerned about the inaccuracy of the tool. One participant found the names of certain information types to be unclear, asking questions like \it{"Expected behavior? Motivation. What's the difference between them?"} - P04. This participant felt that users needed sufficient explanations. This provides us with a suggestion to include the definition of information types in the tool's usage guide. Another argued, usually assigned developers are well aware of the issue discussion. In that case, issue summarization might not help them, rather can help the ones who want to take an overview. \

\subsubsection{Suggestions for Improvements}\leavevmode
\noindent{75\% \it{Participants suggest to introduce and integrate more information }~\color{black!90}\rule{45pt}{6pt}} \\
{25\% \it{Participants suggest to expand for other platform} ~\color{black!50}\rule{15pt}{6pt}} \\
We put some of the suggestions of first two participants into practice, as follows: \begin{inparaenum}[(1)] \item first, in an issue report, we incorporated the information type summary generator that P01 suggested and then we conducted the interview with P02 using the revised tool \item customisation options were added in response to P02's request and the tool was then finalized for the following eight interviews \end{inparaenum}. Here we have listed only the suggestions from these eight participants. Six participants reiterated the desire for additional information, consistent with their earlier feedback about the tool's shortcomings. They wanted features like keyword highlights and timelines to be incorporated, reflecting the recurring need for comprehensive data retrieval. Two participants advocated for expanding the tool to support other ITS, such as the Unity forum, by stating, \it{"I hope there is one in for Unity."} - P04.

\begin{tcolorbox}[flushleft upper, boxrule=1pt, arc=0pt, left=0pt, right=0pt, top=0pt, bottom=0pt, colback=white, after=\ignorespacesafterend\par\noindent]
\textbf{Summary of RQ5.}  Participants found iSum helpful for summarising issues and projects and believe it might benefit developers, project managers, and anybody else new to the project. Most participants expressed their willingness to use this tool. Participants claim that this tool's key advantage is its ability to save time when looking up necessary information and provide a brief summary of a project or problem. Lastly, nine out of ten individuals said they would like to use the tool. The participants suggest that iSum can be more effective if the detection performance of the information types can be further improved. Also, they suggested to enhance the tool for different ITS and asked for more information retrieval from the issue discussion, besides showing the information types. 
\end{tcolorbox}

\section{Threats to Validity}\label{sec:threats}
\textbf{Internal validity} threats refer to the authors' biases while conducting a study. One threat to internal validity is inherent in the participant selection bias of our user studies. We used personal connections to recruit participants, which may not fully represent the broader developer population. Additionally, the scoring criterion in the effectiveness study — where a response is deemed correct if it identifies any one of multiple possible correct answers — is a lenient criterion that may lead to optimistic effectiveness results. However, we argue this reflects real-world issue understanding in scenarios where identifying even one correct answer is practically useful. We acknowledge this as a limitation and recommend a stricter evaluation criterion in future work.
A further threat to internal validity concerns our evaluation protocol: 
sentence-level StratifiedKFold does not guarantee issue-level disjointness. We quantified 
this cross-fold overlap for our built dataset (D2): 10 of the 4,241 test sentences (0.24\%) appear 
verbatim here. On inspection, these are within-issue 
repeated or quoted fragments, for example a command or a quoted benchmark line that recurs 
later in the same thread. Excluding these sentences from evaluation leaves performance 
essentially unchanged: the change is less than 0.01 for every model we evaluated. 
Verbatim cross-fold overlap therefore does not inflate our reported results. We further 
acknowledge that keeping whole issues together during splitting would remove this overlap 
entirely, and that some residual issue-level overlap is not fully captured by a text-level 
analysis.

\textbf{Construct validity} threats refer to errors in the data collection process. We used the same dataset developed and used by Arya et al.~\cite{InfoTypesOSSDiscussion} that was prepared through exhaustive manual labeling by the authors. To create Dataset 2, hours were spent manually annotating the selected issue discussions of the OpenJ9 project repository following the same guidelines used for building Dataset 1, with the first author strictly adhering to the label definitions established by Arya et al \cite{InfoTypesOSSDiscussion}. We chose the most commented closed issues and used the same definition of information types for labeling. However, selecting only the most commented closed issues may not fully represent the entire distribution of issue reports, as shorter or still-open threads may contain different distributions of information types. Furthermore, Dataset 2 was annotated by a single author without independent double-coding or inter-rater agreement measures (e.g., Cohen's $\kappa$). Although the annotation process was frequently reviewed and discussed with all co-authors, including two industry partners from IBM, the absence of formal inter-rater agreement is a limitation. Due to the first author's personal and professional experience bias, interpretation of the issue comments of Dataset 2 may differ in some ways from Dataset 1. The open coding of the feasibility interviews was conducted individually by the first author, which may introduce subjective bias, though all codes and themes were subsequently reviewed and refined collaboratively with co-authors. Additionally, sentences with multiple labels were removed to maintain a single-label classification problem, which may exclude some of the most information-dense content and reduce alignment with real-world issue discussions. To quantify the severity of this exclusion, we recovered the 293 multi-label sentences (6.3\% of Dataset 1) that were removed during dataset construction. Because these sentences already carry expert multi-label annotations in the source corpus, no new annotation was required. We evaluated iSum's LightGBM classifier on these sentences under a leave-one-issue-out protocol, so that each sentence is classified by a model that never observed its issue. 10 of the 293 sentences also carry a rare label that was removed during construction; we set these aside and report the remaining 283 sentences whose labels all fall within the modeled set. Of these 283 multi-label sentences, iSum detected all information types in only 9 (3\%) and at least one type in 97 (34\%), missing at least one type in 274 (97\%). This confirms that excluding multi-label sentences is a genuine limitation for information-dense content, and that a simple sentence-splitting workaround would not recover the full label set. We did not repeat this analysis on Dataset 2 because multi-label sentences are negligible there (26 sentences, 0.6\% of Dataset 2). Additionally, BERT and RoBERTa classify each sentence in isolation, which is an imperfect operationalization of information type for discourse-dependent sentences. We quantified this with an error analysis using ±1 sentence context under issue-grouped 5-fold cross-validation: it yields a small net improvement on Dataset 1 (weighted F1 0.325 to 0.361) but a net decline on Dataset 2 (0.480 to 0.398), helping discussion-oriented types while hurting short, self-contained ones. This confirms single-sentence classification as a genuine but selective construct-validity limitation rather than a uniform one.

\textbf{External validity} threats concern the generalizability of the findings of this study. While more responses would have offered more proof of generalizability, we note that we already observed saturation in our manual coding of themes and labels during open coding. Our effectiveness study considered only one bug report and one feature request from TensorFlow, which limits generalization to other kinds of bugs, features, projects, and issue structures. Furthermore, user study participants were predominantly early-career developers, with only one participant having over five years of experience, which limits our ability to generalize iSum's effectiveness to highly experienced users. We added a cross-dataset zero-shot evaluation, with no target-project fine-
tuning, using our actual models (BERT, RoBERTa, LightGBM, and Random
Forest). Each model is trained on one dataset and tested on the other, over
the 11 labels shared between Dataset 1 and Dataset 2, and we report weighted
F1. Training on Dataset 1 and testing on Dataset 2 (OpenJ9), BERT reaches
0.414, RoBERTa 0.431, LightGBM 0.238, and Random Forest (conversational
features only) 0.202; training on Dataset 2 and testing on Dataset 1 gives 0.435,
0.444, 0.326, and 0.271, respectively. An ensemble of BERT and RoBERTa
improves on either transformer alone in both directions (0.434 and 0.465),
whereas adding LightGBM and Random Forest to that ensemble lowers the
average rather than raising it. These results show that the transformer models generalize substantially
better across datasets than the shallow models, whose reliance on dataset-
specific conversational and lexical patterns does not transfer to an unseen
project. iSum therefore drops moderately on a fully unseen project in the
zero-shot setting, but retains meaningful cross-project signal through its trans-
former components.

\section{Related Work}\label{sec:related-work}
Related work to our study can broadly be divided into three types: \begin{inparaenum}
\item Information overload in SE,
\item Issue report summarization, and
\item Issue discussion exploration\end{inparaenum}. We briefly discuss some of the related works in each category:

\subsection{Information Overload in SE}

Information overload refers to the phenomenon where stakeholders are exposed to an excess of information, much of which is irrelevant to their current tasks. This critical challenge often affects stakeholders' productivity and decision-making capabilities \cite{Kashada2020}. Developers and other project stakeholders are overwhelmed by the sheer volume of information from documentation, version control systems, and bug reports. As software systems grow in complexity, the volume of data that stakeholders must process and manage also increases exponentially. Studies such as those by Meyer et al. \cite{meyer2014software} examine how developers manage and prioritize incoming information from various sources like documentation, version control systems, and bug reports. Their findings suggest that developers often experience cognitive overload, which can reduce the efficiency of debugging and coding activities. As new technologies emerge, software engineers must continuously learn new programming languages, frameworks, and tools. Vasilescu et al. \cite{Vasilescu2015QualityAP} investigated how the adoption of multiple tools and platforms leads to information overload, which can delay decision-making and introduce errors in software development. Research by Bjørn and Ngwenyama \cite{Bjrn2009VirtualTC} looks into how distributed teams handle the coordination of information and the overload it can create, often leading to miscommunication and project delays. Pennington et al. \cite{pennington2007effects} conducted a study to investigate how information overload can impact risk management decisions in software projects. Several software tools have been developed to alleviate information overload by filtering information based on context. One such tool is Mylar \cite{programminginfooverload}, which integrates with the Eclipse IDE and focuses on filtering irrelevant data using task context. Semantic knowledge management systems, explored by Demirsoy et al. \cite{Demirsoy2013UsingSK}, can also address information overload by structuring and categorizing software knowledge.

Modern ITS often manage thousands of tickets per project, making it challenging for developers to prioritize and comprehend the most critical information. Ko et al. \cite{KoSeekDev} investigated how developers navigate bug databases and found that the sheer volume of entries often overwhelms them, leading to inefficiencies in addressing issues.  Anvik et al. \cite{AnvikOpenBugRepo} highlighted how duplicate and redundant bug reports clutter the system, reducing the efficiency of bug triage. They recommended automated techniques to identify and merge duplicate reports. Baysal et al. \cite{infoOverloadITS} explored information overload in issue tracking systems (ITS), highlighting how developers often experience frustration due to the difficulty in accessing necessary information for specific tasks. They point out that the relentless influx of new data, most of which is irrelevant to the immediate needs of developers, complicates the situation further. This excess of non-essential information can lead to critical issues being overlooked. To overcome this, they propose a developer-centric approach to issue tracking, which employs a grounded theory analysis of developer comments to design customized, task-specific dashboards. We also aim to alleviate information overload challenges in ITS by automating the detection and extraction of relevant information from issue discussions and aiding stakeholders in understanding issue reports efficiently and effectively.

\subsection{Issue Discussion Exploration}
Rath and Mäder \cite{RequestforComments} discovered three patterns of conversation in issue threads between stakeholders:  \begin{inparaenum}[(1)] \item monologue \item feedback  and \item collaboration\end{inparaenum}. All of these conversation may contain important information to help developers in their issue resolution process and managers to keep track of the issue. Several works have been done to investigate the issue discussions to retrieve focused information. Viviani et al. \cite{DesignTopicDevDiscuss} analyzed pull request discussion to retrieve design related information. Agrawal et al. \cite{VisualContentOSSIssueJupyter} investigated types of visual contents present in issue discussions.  Sillito et al. \cite{QuesDeveloperSWEvolution} examined developer queries, while  Breu et al.\cite{FreqQuesBugReport} investigated potential impact of the developer-posted queries on issue resolution procedure and Huang et al. \cite{quesinIssueResolving} created a model to forecast the likelihood of questions being raised in issue reports. Significant body of research are dedicated to sentiment analysis \cite{SentimentAnalysisCommitComments,EmotionInIssueCommentOSS,DevEmotionSE,MiningComPatternSWDev,SentimentGitRepo,impactSentimentTonesinIssueDisc} of issue discussion, track down uncivil \cite{IncivilityInOSS} and toxic \cite{Didyoumissmycomment} comments or  presence of human values \cite{HumanValuesInSWDev}. Instead of retrieving a single type of information from the issue discussion, our work tries to capture different types of information to support issue exploration. Multiple works are also present in the literature to support multiple information retrieval from issue discussion. Li et al. \cite{CodeReviewPullBased,ClassificationOfReviewComment} categorized review comments and introduced classification algorithms to better understand the code review process. Arya et al. \cite{InfoTypesOSSDiscussion} uncovered 16 information types from issue discussion and later employed shallow ML models for detection. Paing et al. \cite{QuerTCI} integrated Arya et al.'s classification model \cite{InfoTypesOSSDiscussion} to create a tool that can assist Empirical Software Engineering (ESE) researchers in finding issues containing query keywords. While existing research has made significant advancements, our study focuses on offering better classifying models that surpass SOTA tools \cite{InfoTypesOSSDiscussion} for the detection of information types and developing a framework for quick issue exploration to support developers and project managers. We also conduct an evaluation study to assess the effectiveness of our framework in real-world settings. By incorporating feedback from actual users through this evaluation, we aim to ensure that our tools are not only technologically advanced but also practically useful, aligning with the detailed and diverse needs of OSS stakeholders. 

\subsection{Issue Report Summarization}

ITSs are proven to be rich sources of diverse information about software projects which can be utilized in different software activities. However, these issue reports can be extensive and verbose, leading to information overload. This overload makes it challenging for stakeholders to quickly analyze and extract relevant information. To help developers, testers, and project leads sift through issue reports efficiently, several studies have focused on automated summarization techniques to provide concise summaries of report contents \cite{bugsummarynlp,Krasniqi2021ExtractiveSO}. Rastkar et al. \cite{autoSummarizationOfBugReport} proposed an approach that utilizes NLP and machine learning methods to generate concise summaries of bug reports. Their work demonstrated that machine learning models trained on annotated datasets could effectively identify the most relevant parts of an issue report. Li et al. \cite{Li2018UnsupervisedDB}  proposed DeepSum utilizing unsupervised deep neural networks, while Jiang et al. \cite{crowdSummary} introduced a supervised algorithm called Logistic Regression with Crowdsourced Attributes (LRCA). Jindal et al. \cite{keywordsummary} explored techniques for extracting meaningful keywords and key-phrases to generate summaries with relevant scores. Mukhtar \& Lee \cite{MukhtarLee2023} combine classification and summarization techniques to automatically generate concise summaries of bug reports. Practical AI tools such as Jira AI Summary and Issue Summarizer~\cite{JiraAISummaryRW} demonstrate how AI can summarize ticket descriptions and summaries in Jira workflows. Kogents Jira AI Integration~\cite{KogentsJiraAI} discusses integrating AI into Jira platforms for ticket classification and issue management, while the Eesel AI Blog~\cite{JiraAIEesel} explores best use cases of AI for automating ticket categorization and summarization in software teams. However, the same summary cannot be applicable for all tasks and meet the diverse requirements of the project stakeholders. Therefore, to serve the multifaceted issue summarization and contribute to it, we focused on investigating issue discussion threads and information identification.

Recent works also focus on automating bug triage, classification, and workflow support as part of summarization for stakeholder efficiency. Kiashemshaki et al.~\cite{Kiashemshaki2025} propose an instruction-tuned LLM for automating bug triage with strong shortlist quality. Zhao et al.~\cite{Zhao2025TriageReview} review 234 triage papers, analyzing concepts, architectures, datasets, metrics, and obstacles. Zhang et al.~\cite{Zhang2025BTAL} introduce BTAL, a BERT-TextCNN approach with an adaptive loss to handle imbalance. Andrade et al.~\cite{AndradeClassificationofBugReport} conducted an empirical study evaluating multiple traditional classifiers including SVM for classifying bug reports as bugs, feature requests, or mistakes. Jyothsna and Prasad~\cite{challa2025} proposed an ensemble‑based approach combining SVM, Random Forest, and Logistic Regression for classifying bug reports by their nature (e.g., GUI, performance, configuration), demonstrating the effectiveness of ensemble methods for bug report classification. However, their work operates at the report level rather than classifying individual sentences into fine‑grained information types, as in our work. Nevertheless, their findings reinforce the value of ensemble methods in software engineering classification tasks, which is consistent with our results showing that heterogeneous ensembles outperform stand‑alone models. As an additional stand-alone baseline, we also implemented a linear-kernel SVM, motivated by their observation that the linear kernel outperforms the Radial Basis Function (RBF) kernel for issue report classification. We tuned $C \in \{1, 10, 100, 1000\}$ with \texttt{class\_weight = balanced}, under the same stratified 10-fold protocol, features, and evaluation as our main models. SVM reached a weighted F1 of only 0.42 on Dataset~1 and 0.47 on Dataset~2 (versus 0.61/0.55 for Random Forest and 0.67/0.63 for LightGBM), as its linear boundary struggles with our fine-grained, imbalanced sentence-level task. Given this, we did not include SVM in the main evaluation tables and provide full results in the supplementary material.

\section{Future Work}\label{sec:futurework}

Our current BERT and RoBERTa models classify individual sentences without surrounding context.  Because naive multi-sentence context can help some information types but hurt others, future work should condition context use on information type rather than uniformly widening the input window. Also, we could explore integrating conversational features directly into BERT/RoBERTa inputs via custom fusion layers. Replacing TF-IDF with BM25 \cite{robertson2009probabilistic} or hybrid sparse+dense representations in LightGBM is a promising direction, particularly for handling specialized software engineering vocabularies. We also plan to explore alternative encodings for ensemble predictions and a simpler hybrid approach of feeding RoBERTa predictions directly into LightGBM as an alternative to the full ensemble.

Regarding model selection, we plan to explore code-aware encoders such as CodeBERT \cite{feng2020codebert} and modern encoders like ModernBERT \cite{warner2024modernbert}, which may better handle code-heavy issue discussions. TabPFN \cite{hollmann2022tabpfn} is a promising alternative to LightGBM for our tabular feature space. Instruction-tuned LLMs could be evaluated both as end-to-end classifiers for information type detection and as additional ensemble members or dedicated summarization components, subject to deployment constraints such as latency, cost, and data governance. iSum currently uses TextRank for extractive summarization, and future work could explore more modern summarization techniques, including abstractive and hybrid approaches. We also plan to extend iSum's summaries to include similar related issues and offer deep-dive views of issue reports by information type, showing other relevant comments within each type.

Multi-project leave-one-out cross-validation, where entire projects are held out for evaluation, would provide stronger evidence of cross-project generalization. We also plan to evaluate iSum on a broader distribution of issues, including shorter and still-open threads, to address the current bias towards highly commented closed issues. Sub-sentence labeling is a promising direction to capture multi-label information-dense sentences currently excluded from our datasets. A more rigorous taxonomy refinement via clustering could also produce more orthogonal information types, reducing overlaps between categories such as Investigation and Exploration and Solution Discussion.

The information type labels and BERT/RoBERTa embeddings from iSum could serve as semantic anchors for cross-issue search and retrieval tasks. For example, a developer searching for how a similar bug was reproduced in past issues could query specifically over \textit{Bug Reproduction} classified comments, improving precision over full-text search. Dense retrieval pipelines leveraging our model embeddings could also surface related or duplicate issues more effectively. iSum's 14 conversational features and 16 information types also enable rich analytics for optimizing issue writing and management. Future extensions could compute correlations between information type features and issue resolution time - for example, correlating SUM(\textit{Bug Reproduction} LEN) with issue open time to reveal how detailed reproduction steps impact resolution speed, or correlating SUM(\textit{Expected Behaviour} LEN) with SUM(\textit{Action on Issue} LEN) to highlight requirement-specification mismatches that lead to delays. The existing Manager View could be extended with resolution-time regressions or prescriptive alerts.

For annotation and study methodology improvements, Dataset 2 should be re-annotated involving multiple annotators with adjudication and inter-rater agreement reported (e.g., Cohen's $\kappa$). The feasibility interview open coding should be independently double-coded in future replications. Collecting separate feasibility feedback for bug reports and feature requests would provide more targeted insights, given iSum's stronger performance on bug reports. A stricter correctness criterion requiring identification of all expected answers should also be evaluated alongside the current lenient criterion

Finally, future evaluations should recruit a more balanced pool of participants across experience levels and evaluate iSum on more issue reports across multiple projects. A Likert-scale rating of information type importance would provide more fine-grained insights than the current binary selection. Understanding how iSum integrates with and complements existing developer workflows and toolchains - such as git commit messages, pull request discussions, and project management dashboards - is an important direction. We also plan to expand iSum to support other issue tracking systems beyond GitHub, such as JIRA, as suggested by participants.

\section{Conclusions}\label{sec:conclusions}

Issue report discussion provides a variety of information that may be useful to developers, managers and team leads, yet exploring these detailed discussions can be time-consuming and cumbersome. Recognizing the challenges stakeholders face with navigating extensive issue discussions, we designed a tool named iSum that provides issue and project summarization using information types found in issue discussions. The effectiveness of iSum was first established through performance improvements over SOTA \cite{InfoTypesOSSDiscussion} tools in accurately detecting various information types within issue discussions. Subsequently, we conducted a user study involving tasks that required efficient understanding of two types of issue reports—bug and feature—both with and without the assistance of iSum. The results from this study highlighted that iSum supports more effective and efficient understanding of bug reports between the two, significantly enhancing stakeholders' ability to understand issues within a short time frame. Our future work will focus on extending iSum's capabilities by incorporating richer summary views, improving classification performance for feature requests, and evaluating iSum across a broader range of projects and issue types to further validate its generalizability.

\section{Declarations}
\textbf{Data Availability.} The code and data used for this paper are shared in this repository: \url{https://github.com/tamanna037/ISum}\\
\textbf{Conflict of Interest.} The authors have no conflict of interest.\\
\textbf{Ethics Approval.} This study was reviewed and received ethics clearance from the Conjoint Faculties Research Ethics Board of 
the University of Calgary.\\
\textbf{Informed Consent}
All participants provided verbal informed consent prior to participation in the study. \\
\textbf{Clinical Trial Number}
Not applicable.\\
\textbf{Funding.} This study was supported by IBM Canada. \\
\textbf{Author Contributions} 
S.B. Tamanna contributed to the investigation, data curation, methodology design, development, and preparation of the manuscript.
J.Y. Khan assisted with certain aspects of the ML model development.
G. Uddin supervised the study and contributed to manuscript review and editing.
L. Xia and L. Zhang contributed to project administration and manuscript review.
All authors have read and approved the final version of the manuscript.

\urlstyle{rm}
\bibliographystyle{abbrv}
\bibliography{References}
\newpage

\end{document}